\documentclass{iopart}

\usepackage{graphicx}
\usepackage{setstack}
\usepackage{stmaryrd}
\usepackage{amssymb}
\usepackage{amsbsy}
\usepackage{amsfonts}
\usepackage{mathrsfs}
\usepackage{tikz}
\usepackage{pst-all,pstricks}
\usepackage{bbold}
\usepackage{upgreek}
\usepackage{bm}
\usepackage{arydshln}
\usepackage{subfig}
\usepackage{rotating}
\usepackage{color}
\usepackage{backref}
\renewcommand*{\backref}[1]{}
\renewcommand*{\backrefalt}[4]{%
\ifcase #1 %
\or
$\quad$({\footnotesize\it Cited p.~#2})%
\else
$\quad$({\footnotesize\it Cited p.~#2})%
\fi
}

\newtheorem{theorem}{Theorem}
\newtheorem{proposition}[theorem]{Proposition}
\newcommand{\vect}[1]{\mathbf{#1}}
\newrgbcolor{cafe}{.627 .25 0}
   \definecolor{USgray}{rgb}{0.5469,0.5895,0.6016}
\newrgbcolor{Cafe}{.447 .294 .027}
\newrgbcolor{Verde}{.1 .447 .027}
\def\Tiny{\fontsize{1pt}{1pt}\selectfont}

\newlength{\myL}

\newcommand\hexagonoc[1]{%
  \setlength{\myL}{#1}
  \psset{unit=1.0}
\begin{rotate}{-30}
  \begin{pspicture}(-0.9511\myL,-0.8090\myL)(0.9511\myL,\myL)
  \psline[linewidth=1pt](-0.5\myL,-0.866\myL)(-\myL,0)(-0.5\myL,0.866\myL)%
  (0.5\myL,0.866\myL)(\myL,0)(0.5\myL,-0.866\myL)
  \psline[linewidth=1pt,linestyle=dashed,dash=.5mm .5mm .5mm .5mm](0.5\myL,-0.866\myL)(-0.5\myL,-0.866\myL)
  \psdots[dotstyle=o,dotsize=5pt](0.5\myL,-0.866\myL)(-0.5\myL,-0.866\myL)(-\myL,0)(-0.5\myL,0.866\myL)(0.5\myL,0.866\myL)(\myL,0)
  \end{pspicture}\end{rotate}
  \rput{0}(2.3\myL,-0.35\myL){$\quad u(1)^2$}
  \rput{0}(-.35\myL,-0.35\myL){$u(1)^{n-2}$}
  \rput{0}(2.3\myL,0.75\myL){$\quad u(1)$}
  \rput{0}(-.35\myL,0.75\myL){$u(1)^{n-1}$}
  \rput{0}(1.125\myL,-1.1\myL){$u(1)^3$}
  \rput{0}(1.125\myL,1.5\myL){$u(1)^n$}
}

\makeatletter
\def\underbracket{%
\@ifnextchar[{\@underbracket}{\@underbracket [\@bracketheight]}%
}
\def\@underbracket[#1]{%
\@ifnextchar[{\@under@bracket[#1]}{\@under@bracket[#1][0.2em]}%
}
\def\@under@bracket[#1][#2]#3{
\mathop{\vtop{\m@th \ialign {##\crcr $\hfil \displaystyle {#3}\hfil $%
 \crcr \noalign {\kern 3\p@ \nointerlineskip }\upbracketfill {#1}{#2}
 \crcr \noalign {\kern 3\p@ }}}}\limits}
 \def\upbracketfill#1#2{$\m@th \setbox \z@ \hbox {$\braceld$}
 \edef\@bracketheight{\the\ht\z@}\bracketend{#1}{#2}
 \leaders \vrule \@height #1 \@depth \z@ \hfill
\leaders \vrule \@height #1 \@depth \z@ \hfill \bracketend {#1}{#2}$}
 \def\bracketend#1#2{\vrule height #2 width #1\relax}
\makeatother

\begin{document}

\title{Magnetic operations: a little fuzzy mechanism?}
\author[cinves]{B. Mielnik$^1$ and A. Ram\'irez$^2$.} 
\address{$^1$ Departamento de F\'isica, Centro de Investigaci\'on y de Estudios Avanzados del I.P.N., Apdo. Postal 14-740, M\'exico D.F. 07000, M\'exico.}
\address{$^2$ Soporte a Infraestructura de Redes LAN, Direcci\'on de Inform\'atica,  Radiom\'ovil DIPSA S.A. de C.V.,  Lago Zurich 245. M\'exico D.F. 11529, M\'exico. }
\eads{\mailto{bogdan@fis.cinvestav.mx}, \mailto{alejandra.palacios@mail.telcel.com}}

\begin{abstract}
We examine the behaviour of charged particles in homogeneous, constant and/or oscillating magnetic fields in the non-relativistic approximation. A special role of the geometric center of the particle trajectory is elucidated. In quantum case it becomes a `fuzzy point' with non-commuting coordinates, an element of non-commutative geometry which enters into the traditional control problems. We show that its application extends beyond the usually considered time independent magnetic fields of the quantum Hall effect. Some simple cases of magnetic control by oscillating fields lead to the stability maps differing from the traditional Strutt diagram.
\end{abstract}



\section{Introduction}
\label{sec1}
The present day quantum theories offer some visions of new, mathematically possible reality which, until now, were not experimentally detected. Thus, the idea of supersymmetry looks for particles in superposed states of `being fermion' and `being boson' and for the corresponding invariance group, with consequences for elementary particles, strings, and quantum cosmology  \cite{Higgs:1964,Higgs:1966,Kane:2000,Misner:1972js,PhysRevD.40.3982,Encyclo}. The other models propose to avoid the field singularities by assuming that the very structure of the physical space forbids the exact point localizations.  In some of them (Connes, Madore, Bellisard) the physical points have the non-commuting coordinates \cite{Connes:1994, Madore:1992, Seiberg:1999vs, Connes:2004,Majid:1999td,Balachandran:2002,Mad,Belli,Oconnor:2003,MR2287303,GarciaCompean:2001wy,GarciaCompean:2002ss,LopezDominguez:2006wd}. According to Doplicher et al \cite{Doplicher:1994,Doplicher:1995}, this might  explain why the exact position measurements in the Heisenberg's microscope cannot produce the microscopic black holes, (even if the reality of the danger is an open problem). In the simplest model of the non-commutative plane (Madore \cite{Madore:1992}) the points $(x,y)$  fulfill $[x,y]=i\kappa\neq 0$, preventing as well the singularity creation.\\

Though all these ideas are just free associations (in almost psychoanalytic sense), no better justified is the traditional hypothesis that our space is indeed a continuum of the `exact points'. In fact, if adopting Wigner's observation about the ``unreasonable power of mathematics" \cite{Wigner:1979}, the mathematical models, if correct, must appear somewhere in nature. So indeed happens, but with one amendment.  The nature might `use' our models in its own way, without caring about the author's intentions. Thus, while the problem of boson-fermion supersymmetry waits to be solved (still no trace of Higgs boson, no gravitinos, etc.), the same mathematical structure appeared in the supersymmetric quantum mechanics (Witten \cite{Witten:1981,Witten:1982}) helping to solve exactly a class of spectral problems (Duplij et al \cite{Encyclo}).  While there is still no sign of strings, branes nor extra dimensions, the analogous mathematical structures permit to understand better certain biological phenomena. In fact, the 'clones' of almost all unborn structures start invading the present day physics forming a little-great science in its own right (c.f. ``big-bang in probet'' etc. \cite{Ashtekar:2006,Zurek:1996}).\\

Below, we shall discuss a similar status of the `non-co\-mmu\-ta\-ti\-ve points' which appear in the quantum Hall effect. We shall show that the idea works not only for the static fields but also for time dependent ones, even though the more fundamental hypothesis concerning the granular structure of the space itself is still waiting to be confirmed (or abandoned?). 

\section{The charged particles in homogeneous magnetic fields: circular and drifting trajectories.}\label{sec2}
We start from the known facts.
The time independent magnetic fields $\vect{B}$ in an open domain of $\mathbb{R}^3$ can be described by a class of the vector potentials $\vect{A}(\vect{x})$ with $\vect{B}(\vect{x})=rot\vect{A}=\vect{\nabla}\times\vect{A}(\vect{x})$. If $\vect{B}(x)$ is homogeneous, $\vect{B}=(0,0,B)$ (for convenience, let $\vect{B}$ define the $z$-axis in $\mathbb{R}^3$) then one of natural choices of $\vect{A}(\vect{x})$ is:
\begin{eqnarray}\label{potencialvectorial}
\vect{A}(\vect{x}) = \frac{1}{2}\vect{B}\times\vect{x}=\frac{1}{2}B\left\|\begin{array}{c}
-y \\[-2pt] x \\[-2pt]0
\end{array}\right\|,
\end{eqnarray} 
interpretable as the vector potential created by a homogeneous current density on the cylindrical surface. The particle motion along the $z$-direction is then free; hence, we shall be interested only in the motion trajectories on the $x,y$-plane. In the non-relativistic approximation, the Hamiltonian of a classical point particle of charge $e$ and mass $m$ is
\begin{eqnarray}\label{hamiltonianocompleto}
 H_\beta = \frac{1}{2m}\left(\vect{p}-\frac{e}{c}\vect{A}\right)^2 = \frac{1}{2m}\left[(p_x+\beta y)^2+(p_y-\beta x)^2\right]
\end{eqnarray}
or equivalently:
\begin{eqnarray}\label{hamiltoniano}
 H_\beta = \frac{1}{2m}\left[\vect{p}^2+\beta^2\vect{x}^2\right]-(\beta/m)M_z
\end{eqnarray}
where $\beta=\frac{eB}{2c}$; $\vect{x}=(x,y)$ and $\vect{p}=(p_x,p_y)$ are the generalized momenta, and $M_z=xp_y-yp_x$. The well known shape of the motion trajectories is most easily derived from the Hamiltonian in form (\ref{hamiltonianocompleto}). The first pair of the canonical eqs. defines the interrelation between the generalized and kinetic momenta $m\vect{v}$
\begin{eqnarray}\label{momentos}
\eqalign{\frac{dx}{dt}=\frac{\partial H_\beta}{\partial p_x} = \frac{1}{m}(p_x+\beta y) &\qquad \Rightarrow \qquad
 p_x = mv_x-\beta y\\
\frac{dy}{dt}=\frac{\partial H_\beta}{\partial p_y} = \frac{1}{m}(p_y-\beta x) &\qquad \Rightarrow \qquad
 p_y = mv_y+\beta x}
\end{eqnarray}
and the second pair yields the proper dynamical equations:
\begin{eqnarray}\label{ecsdinamicas}
\eqalign{\frac{dp_x}{dt}=-\frac{\partial H_\beta}{\partial x} &= \frac{\beta}{m}(p_y-\beta x)\\
\frac{dp_y}{dt}=-\frac{\partial H_\beta}{\partial y} &= -\frac{\beta}{m}(p_x+\beta y)
}
\end{eqnarray}
Both (\ref{momentos})-(\ref{ecsdinamicas}) imply immediately the existence of two conservative quantities:
\begin{eqnarray}
 X = \frac{x}{2}+\frac{p_y}{2\beta} = x+\frac{mv_y}{2\beta} = x+\frac{v_y}{\omega},\label{conservativas}
\end{eqnarray}
\begin{eqnarray}
 Y = \frac{y}{2}-\frac{p_x}{2\beta} = y-\frac{mv_x}{2\beta} = y-\frac{v_x}{\omega}.\label{XY}
\end{eqnarray}
with $\frac{d}{dt}X=\frac{d}{dt}Y=0$, hence
\begin{eqnarray}\label{orbitas}
 \frac{d}{dt}\left\|\begin{array}{c}
              x-X\\y-Y
             \end{array}\right\| = \frac{2\beta}{m}\left\|\begin{array}{c}
                                             y-Y\\-(x-X)
                                           \end{array}\right\|
\end{eqnarray}
so, each charged particle just rotates around a fixed center $\vect{X}=(X,Y)$ with a constant (cyclotron) frequency $\omega = \frac{2\beta}{m} = \frac{eB}{mc}$:
\begin{eqnarray}\label{trayectorias1}
 \left\|\begin{array}{c}
  x(t)-X\\y(t)-Y
\end{array}\right\| = \left\|\begin{array}{rc}
                 \cos(\omega t) & \sin(\omega t)\\
                -\sin(\omega t) & \cos(\omega t)
               \end{array}\right\|\left\|\begin{array}{c}
                             x(0)-X\\y(0)-Y
                            \end{array}\right\| .
\end{eqnarray}
The conservative quantities $X, Y$  represent the hidden symmetries of the system \cite{Connes:2004,Mad,Belli}. The expression for $X,Y$ in terms of the generalized momenta might look peculiar but it turns natural in terms of the velocities (kinetic momenta) (\ref{momentos}). It shows that the radius $\rho$ of each rotating trajectory depends just on its (constant) velocity scalar $v=|\vect{v}|=\sqrt{(v_x^2+v_y^2)}$:
\begin{eqnarray}\label{radio}
 \rho^2=(x-X)^2+(y-Y)^2=\frac{v^2}{\omega^2}. 
\end{eqnarray}
Yet, to find the system response to the external forces, the most convenient are the expressions (\ref{conservativas}-\ref{XY}) in terms of $p_x, p_y$.  Indeed, if the circulating charge  (\ref{hamiltonianocompleto})-(\ref{trayectorias1}) is affected by an additional potential $V(\vect{x})$, the Hamiltonian becomes $\tilde{H}=H_\beta+V(\vect{x})$ and since the motion center is conserved by $H_\beta$, the canonical eqs. for $X, Y$ are reduced to:
\begin{eqnarray}
  \frac{dX}{dt} &= \{X,V\} = \frac{1}{2}\left\lbrace x+\frac{p_y}{\beta},V(x,y)\right\rbrace  = -\frac{1}{2\beta}\frac{\partial V}{\partial y}=\frac{1}{2\beta}F_y\\[10pt]
  \frac{dY}{dt} &= \{Y,V\} = \frac{1}{2\beta}\frac{\partial V}{\partial x}=-\frac{1}{2\beta}F_x \label{canonicas}
\end{eqnarray}
 \begin{figure}[b]%
 \begin{minipage}{2cm}%
 \hspace*{.5cm} \includegraphics[scale=.4]{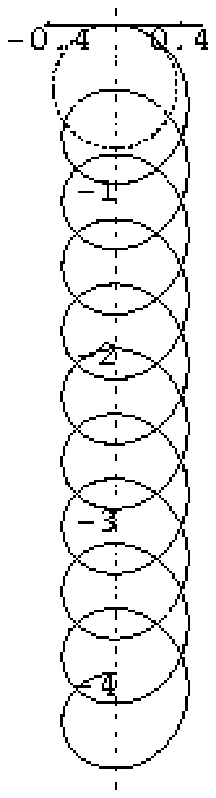}
 \end{minipage}
 \qquad
 \begin{minipage}{5.5cm}%
 \hspace*{.35cm}\includegraphics[scale=.4]{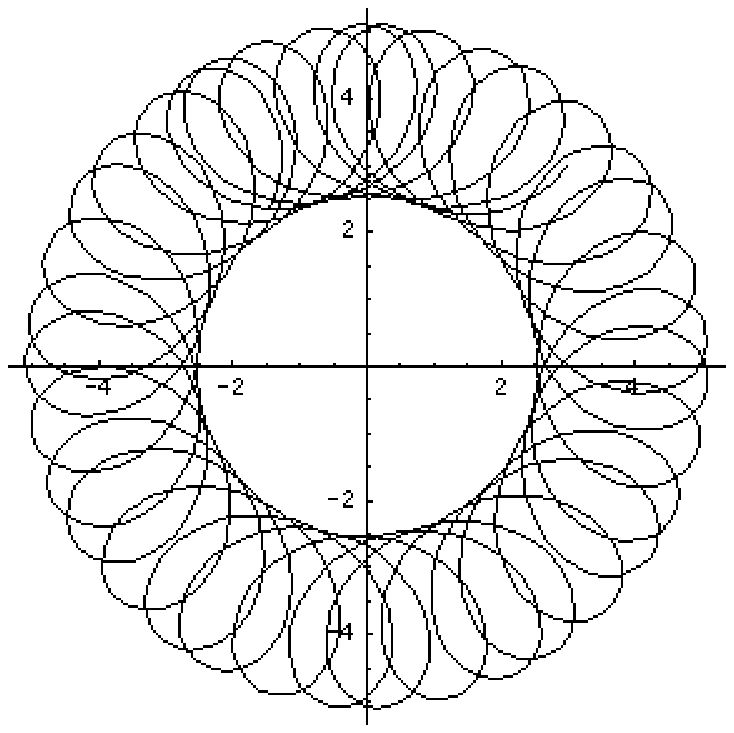}
   \rput(-6.4,3){ $a)$}
  \rput(-4,3){ $b)$}
   \rput(2.2,3){ $c)$}
 \end{minipage}%
 \qquad
 \begin{minipage}{1in}%
 \hspace*{.35cm}\includegraphics[scale=.4]{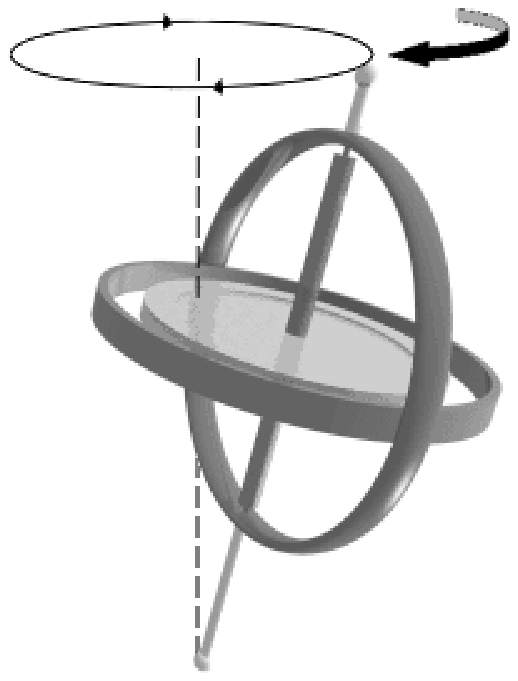}
 \end{minipage}
 \begin{quote}
 \caption{\label{fig1}\small The three generically similar phenomena. a) The rectilinear drift along the axis $y$ under a constant force $F$ in the $x$ direction. b) The circular motion trapped by a repulsive oscillator. c) The precession of a gyroscope.}
 \end{quote}
 \end{figure} 
In the simplest case, if the trajectory (\ref{trayectorias1}) is affected by a constant force $F$, the well known though counter-intuitive effect is that the rotation center starts to drift in the direction orthogonal to $F$. Curiously, the exact solutions of (\ref{canonicas}) exist also for the elastic potentials $V(\vect{x})=\alpha\vect{x}^2$. For $\alpha<0$, $|\alpha|<\frac{\beta^2}{m}$, they illustrate the surprising phenomenon of a charged particle trapped by the repulsive center (Fig. \ref{fig1}). 

A less elementary form of the same effect, in presence of crossed electric and magnetic fields is discussed in \cite{Piasecki:1996}, where the KAM theorem \cite{Poschel:2001} is applied to show that the existence of an additional repellent obstacle (in form of a disk) can interrupt the rectilinear drifting, trapping the charged particle which instead of being rejected, `obsessively returns' to the repelling obstacle. Below, we shall not pretend to deepen this line of thought, but instead, we shall focus attention on the corresponding quantum systems.

\section{Quantum case: the ``fuzzy points''.}
\label{sec3}
\begin{flushright}
 \begin{footnotesize}``\textit{Pure quantum states are objective but not real}''\\[4pt]
\end{footnotesize}
\begin{scriptsize}
Hans Primas,
\textit{Chemistry, Quantum Mechanics and} \\
 \textit{Reductionism.}, Springer Verlag (1983) p. 103.\\[15pt]
\end{scriptsize}\end{flushright}

The quantum equivalents of (\ref{potencialvectorial}-\ref{canonicas}) are so widely studied, that we fix attention just on one particular aspect. We shall use the same symbol $H_\beta$ to denote the quantum Hamiltonian (\ref{hamiltonianocompleto})-(\ref{hamiltoniano}), though now $x_k, p_l$ will mean the quantum observables, $[x_k,p_l]=i\delta_{kl}$. By commuting the Hamiltonian (\ref{hamiltonianocompleto}) with $x_k, p_l$ one obtains the same two pairs of equations (\ref{momentos}) and (\ref{ecsdinamicas}), though now they concern the Heisenberg's operators $x(t), p_x(t), y(t), p_y(t)$. In a complete analogy with the classical case, the formulae (\ref{momentos})-(\ref{ecsdinamicas}) show the existence of a pair of conservative observables $X, Y$ given by (\ref{conservativas}-\ref{XY}). 

The works on quantum systems (\ref{hamiltonianocompleto})-(\ref{hamiltoniano}) dedicate a lot of care to the non-commuting kinetic momenta \cite{Connes:1994,Torres:2006}. It neither escapes attention that the same phenomenon affects the coordinates of some abstract `space localizations' such as the rotation center of the Heisenberg's trajectory (c.f. Avron et al. \cite{Avron:1978}, Dodonov et al. \cite{Dodonov:1994}):
\begin{eqnarray}
 X=\frac{1}{2}x+\frac{1}{2\beta}p_y\label{XyY-X}\\[7pt]
 Y=\frac{1}{2}y-\frac{1}{2\beta}p_x\label{XyY-Y}
\end{eqnarray}
While the instantaneous particle coordinates $x(t), y(t)$ at any fixed time moment $t$ are still the commuting observables, this is no longer true for the rotation center $\vect{X}=(X,Y)$ which becomes a ``fuzzy point'' \cite{Madore:1992}:
\begin{eqnarray}\label{conmutador}
 [X,Y]=-\frac{i}{2\beta}=-\frac{i}{m\omega}
\end{eqnarray}
Note that, the ``fuzzy localization'' is an abstract concept (resembling a ``grin without the cat'' of Lewis Carroll \cite{Carroll:1865}). There is simply nothing there, commuting or not. Yet, by paraphrasing Primas \cite{Primas:1983}, the fuzzy center (\ref{XyY-X})-(\ref{XyY-Y}) is ``\textit{not real but is objective}''.  

Another curious aspect, resembling some open gravity problems \cite{PhysRevD.44.1740,4164,Rovelli:2004} concerns the `surface' of the $2$-dimensional orbits (\ref{trayectorias1},\ref{radio}). Indeed:
\begin{eqnarray}\label{superficie}
\fl \pi\rho^2 = \pi(\vect{x}-\vect{X})^2
= \frac{\pi}{4\beta^2}\left[ \Bigl( p_x+\beta y\Bigr)^2+\Bigl( p_y-\beta x\Bigr)^2\right] =\frac{\pi}{(2\beta)^2}H_\beta
\end{eqnarray}
Hence, the surface of the orbit is not only conserved but also quantized: it is proportional to the Hamiltonian and cannot change continuously (an equivalent observation see, e.g. \cite{Vagner}).

As it seems, this is not the first time when the magnetic fields provide an imitation of still unchecked theories. The suggestive ideas of supersymmetry are still not verified in particle physics. Yet, due to the anomalous relation between the spin and orbital magnetic moments of the electron, its energy levels in a static, homogeneous magnetic field reproduce the supersymmetric spectrum. So, many authors conclude that: ``the supersymmetry exists in nature'' \cite{Gendenshtein:1983,Gendenshtein:1986ub,Bagchi:2000,BMORO}. Quite similarly, there is no evidence that the physical particles move in non-commuting spaces, but the properties of the rotation center (\ref{conmutador}) might encourage the statement that ``the non-commutative positions exist in nature''. 

 It might be thus interesting to notice that the  phenomenon is not limited to the circular trajectories in the static magnetic fields. It appears as well in  more general physical scenarios, including the time dependent oscillator potentials and magnetic fields.


\section{Time dependent oscillators: the non-circular loops}\label{sec4}

The evolution problem for variable oscillator potentials has already a notable past. The first systematic studies were presented in 1967-69 by Lewis and Riesenfeld \cite{Lewis:1967,Lewis:1968}, then by Malkin and Man'ko \cite{Malkin:1970}, inspiring an ample research on the state evolution. The group theoretical approach to the Baker-Campbell-Hausdorff (BCH) problem \cite{Baker:1898,Campbell:1905,Hausdorff:1906,suzuki:601} for quadratic Hamiltonians was developed in 1974 by Gilmore \cite{Gilmore:1974} (a generalization see Zhang et al \cite{Gilmore:1990}). In 1976 Yuen notices the possibility of solving the evolution equation for the squeezed photon states \cite{Yuen:1976}. For massive charged particles the operations induced by variable fields were studied by Ma and Rhodes \cite{Ma:1989}, Royer \cite{Royer:1985}, Brown and Carson \cite{Brown:1979}, Combescure \cite{Combe}, Wolf \cite{Wolf:1981} and other authors; for an ample review c.f. Dodonov \cite{Dodonov:2002}.  The links with the coherent states are notable \cite{Manko:1997,Glau1,Glau,Alonso}. In the parallel research there was hardly any constant in the generalized oscillator or magnetic Hamiltonians which would not be replaced by its time dependent analogue, starting from the time dependent mass ($c.f.$ the Hamiltonian of Caldirola-Kanai \cite{Caldirola:1941,Kanai:1948,Compean:1997}) up to the variable dielectric or magnetic permeability \cite{Castanos:1997,Castanos:1994,Castanos:1996,DKN}.

The possibility of the closed non-circular trajectories for the time dependent oscillator Hamiltonians was noticed already in Malkin and Man'ko \cite{Malkin:1970}. The extremely simple cases of the closed evolution caused by the sudden $\delta(t)$-shocks of the oscillator potentials were described  in \cite{BM77,Bogdan:1986,Bogdan:1994,David:1994}. Thus,  for the Schr\"odinger's particle in 1 space dimension the \textit{evolution loops} can be produced by  sequences of oscillator pulses and free evolution intervals corresponding  to the elementary cases of the $BCH$ formula, e.g.
\begin{eqnarray}\label{12terms}
\underbracket{e^{-i\tau\frac{p^2}{2}}e^{-\frac{3i}{\tau}\frac{x^2}{2}}\cdots e^{-i\tau\frac{p^2}{2}}e^{-\frac{3i}{\tau}\frac{x^2}{2}}}_{6\;\,\mathrm{terms}}\equiv1
\end{eqnarray}
(with $\hbar=1$), illustrated by the evolution diagramme:
\begin{center}
\begin{pspicture}(0,.75)(6,4.5)
    \pcline{o-o}(1,1)(5,1)
    \pcline{o-o}(5,1)(3,4)
    \pcline{o-o}(3,4)(1,1)
\rput{0}(3,1.25){$\tau$}
\rput{0}(2.25,2.45){$\tau$}
\rput{0}(3.75,2.45){$\tau$}
\rput{0}(3,4.5){$\frac{3}{\tau}$}
\rput{0}(.75,1){$\frac{3}{\tau}$}
\rput{0}(5.25,1){$\frac{3}{\tau}$}
\end{pspicture}
\end{center}
where the vertices symbolize the shocks of the elastic potential (with the corresponding numbers, meaning the pulse amplitudes) and the sides correspond to the `rest intervals' of the free evolution (for simplicity we accept the particle mass $m=1$). The equivalence sign $\equiv$ in (\ref{12terms}) means the operator proportionality, $i.e.$, $U\equiv1 \Leftrightarrow U=e^{i\varphi}1$, $ \varphi\in \mathbb{R}$. The many vertex analogues of (\ref{12terms}) as well as the more general 'kicked systems'  and their non-singular equivalents were described in \cite{Bogdan:1986,Bogdan:1994,Dodonov:1993,Harel:1999,Niu:2000,Fish,Fish2,Viola1999,Gerardo,Bogdan:1990,Lin}.

Some incomplete versions of the loop process may prove of interest. In fact, whenever an evolution loop contains a $\delta$-pulse of the attractive oscillator potential, as in (\ref{12terms}), the whole rest of the process must imitate the effect of a repulsive elastic pulse (which cannot be straightforwardly achieved by the magnetic fields). Furthermore, if any evolution loop contains an interval of free evolution $e^{-i\tau\frac{p^2}{2}}$ $(\tau>0)$, it means that the whole rest must be equivalent to its inverse, thus suggesting the techniques of reverting the free propagation. The importance of the closed dynamical processes for more general control operations was recognized in \cite{BM77,David:1994,Harel:1999,Niu:2000,Viola1999}.

All these models depend on some idealizations. Thus, the repulsive oscillator kicks cannot be straightforwardly simulated by the magnetic fields (in fact, even the attractive ones can hardly be engineered!). Moreover, in order to describe the time dependent potentials in some little but ma\-cros\-co\-pic areas as, $e.g.$,  an ion trap, one typically uses the non-relativistic \textit{laboratory approximation}, disregarding the little delays needed to propagate the potential inputs all over the trap surfaces (telegraphist's equations) or in its interior. Since these delays in typical experiments are insignificant, the lab approximation works very well and indeed, is implicit in all papers postulating the time dependent external parameters  ($e.g.$ \cite{Lewis:1967,Malkin:1970,Bogdan:1994,Bogdan:1986,David:1992,Castanos:1997,Manko:1997}). 

Once this approximation is adopted, one can see that the fuzzy centers
are not restricted to the orthodox Hall effect with a fixed magnetic background. They arise, with comparable consequences, for arbitrary evolution loops generated by the static or time dependent quadratic Hamiltonians.


\section{Loops of quadratic Hamiltonians: stability and drifting.}
\label{sec5}

To show this, we shall consider the hermitian Hamiltonians $H(t)$ defined by time dependent \textit{quadratic forms} $H(t)=\sum\limits_{i,j=1}^{2s}h_{ij}(t)q_iq_j$ (with $h_{ij}(t)=h_{ji}(t)\in\mathbb{R}$) of a complete set $q_1,\cdots, q_{2s}$ of any number of the \textit{canonical observables}, $x_i, p_i$ $(i=1,...,s)$ in a certain Hilbert space $\mathscr{H}$; $[x_i,x_j]=[p_i,p_j]=0,$  $[x_i,p_j]=i\delta_{ij}$. An agreeable property of the quadratic Hamiltonians is that even though $H(t)$ are unbounded, but if the coefficients $h_{ij}(t)$ are non-singular and piece-wise continuous, the corresponding unitary evolution operators $U(t,\tau)$ are well defined by the operator equations:
\begin{eqnarray}\label{BogA20}
 \frac{d}{dt}U(t,\tau)=-i\,H(t)U(t,\tau) \;\Leftrightarrow\; \frac{d}{d\tau}U(t,\tau)=i\,U(t,\tau)H(\tau),
\end{eqnarray}
with $U(\tau,\tau)=\mathbf{1}$,  and:
\begin{eqnarray}\label{BogA21}
 U(t,\tau)U(\tau,\sigma)= U(t,\sigma),
\end{eqnarray}
(deeper results and amendments, c.f. B. Simon \cite{Simon:1971}, Hagedorn et al \cite{Hage}). As well known, for quadratic $H(t)$, the  time dependent Heisenberg's observables $x_i(t), p_i(t)$ are linear combinations  of the initial $x_i, p_i$. For convenience, we shall denote by $\vect{q}$ the vector-column of $2s$ (dimensionless) observables:
\begin{eqnarray}
\vect{q}=\left\|\begin{array}{c}\label{q}
 q_1\\ \vdots\\[-10pt]q_{2s}
\end{array}\right\|=\left\|\begin{array}{c}\,
\vect{x}\\\vect{p}\,
\end{array}\right\|,
\end{eqnarray}
One therefore has:
\begin{eqnarray}\label{BogA22}
 \vect{q}(t)=U(t,0)^\dagger\vect{q}U(t,0)=u(t)\vect{q}
\end{eqnarray}
where $u(t)=u(t,0)$ is a dimensionless $2s\times 2s$ \textit{evolution matrix} yielding simultaneously the classical and quantum trajectories.

If $\vect{q}$ generates all observables in $\mathscr{H}$, then not only the evolution operator $U(t)$ determines the matrix $u(t)$, but inversely,  the matrix (\ref{BogA22}) determines the unitary $U(t)$ up to a numerical phase factor. Indeed, should $U$ and ${U'}$ be two unitary operators generating the same transformation of the canonical variables $\vect{q}$, i.e. $U^\dagger\vect{q}U={U'}^\dagger\vect{q}{U'}$, then ${U'}U^\dagger\vect{q}=\vect{q}{U'}U^\dagger$, and so, ${U'}U^\dagger$ would commute with all canonical variables $\vect{x}, \vect{p}$ and their functions. However, if the algebra spanned by the canonical observables $\vect{q}$ is irreducible in  $\mathscr{H}$, then any operator which commutes with all of them must be just a c-number. Since $U$ and ${U'}$ are unitary, this number can only be a  phase factor, i.e.: $ {U'}U^\dagger=e^{i\varphi}$ implying ${U'}=e^{i\varphi} U $. The concrete value of $\varphi$ is relevant for the transformation of the \textit{state vectors} but not of the Heisenberg's variables, so we shall simply denote: 
\begin{eqnarray}\label{UU}
{U'}=e^{i\varphi} U\;\Longrightarrow\;{U'}\equiv U 
\end{eqnarray}
 This equivalence turns quite essential in quantum control, as it basically permits to deduce the form of $U$ from the transformation of the canonical observables. Moreover, since for the quadratic $H(t)$ the evolution matrix in quantum and classical cases is the same, the properties of the evolution operators $U(t)$ can be read as well from the classical motion trajectories. This allows numerical but non-perturbative solutions of the continuous  BCH  problem 
\cite{Magnus:1954,Gilmore:1974,BMP,Bogdan:1970}
 for the evolution (\ref{BogA20}). In particular, an evolution loop occurs if after a certain time interval (for convenience, let it be $[0,T]$) all canonical (classical and/or quantum) variables $\vect{q}$ return to their initial values:
\begin{eqnarray}
 \vect{q}(T)=U(T,0)^\dagger\vect{q}U(T,0)=\vect{q}\quad\iff\quad u(T)=\mathbb{1},
\end{eqnarray}
so that, in the sense of (\ref{UU}):
\begin{eqnarray}\label{UT0}
 U(T,0)=e^{i\varphi}\vect{1}\equiv\vect{1},\qquad (\varphi\in\mathbb{R}),
\end{eqnarray}
implying also a loop of all other observables $A(t)$ which do not depend explicitly on time in the Schr\"odinger's frame, i.e., $A(T)=U(T,0)^\dagger AU(T,0)= A$. The value of $\varphi\in\mathbb{R}$ ($i.e.$, the geometric phase for the loop process) though interesting in itself \cite{Pleban:1989,David:1994}, does not affect the results of our present argument. 

Below, we shall consider a Hamiltonian $H(t)$ varying periodically, i.e., $H(t+T)= H(t)$. We denote for simplicity $U(t,0)= U(T+t,T)=U(t)$. We assume moreover that the evolution produces a loop in $[0,T]$ and in the subsequent periodicity intervals. Then, any Heisenberg's observable $A(t)=U(t)^\dagger AU(t)$ in any periodicity interval $[\tau,\tau+T]$ admits the time average
\begin{eqnarray}\label{Apromedio}
 \widehat{A}=\frac{1}{T}\int\limits_\tau^{\tau+T}A(t)dt,
\end{eqnarray}
independent on $\tau$, defining a certain global characteristic of the loop process. (Indeed, the change of $\tau$ in (\ref{Apromedio}) means  just that the same contributions $A(t)dt$ are summed up along the same closed cycle, rearranging only the summation order)\footnote{This does not mean that $\widehat{A}$ is a  constant of motion in a conventional sense. As an integral over the entire time interval $[\tau, T+\tau]$, the observable $\widehat{A}$ is not `local in time'. Yet, if  the little evolution steps of  $A(t)$  in (\ref{Apromedio}) obey their (different) instantaneous Hamiltonians, then (\ref{Apromedio}) stays unchanged.}.

In particular, applying (\ref{Apromedio}) for the time dependent variables $x_j(t)$, one obtains the coordinates $X_j$ of the `loop center'. For $H(t)$ quadratic they are as well linear in the initial $q_1 , ... , q_{2s}$:
\begin{eqnarray}\label{X}
\fl \vect{X}=\widehat{\vect{x}}\quad\Rightarrow\quad X_j=\widehat{x}_j=\widehat{u}_{j1} q_1+\cdots+\widehat{u}_{j\,2s}\, q_{2s}\qquad\quad (j=1,\cdots,s),
\end{eqnarray}
generalizing the already described rotation centers (\ref{XyY-X})-(\ref{XyY-Y}) in 2D. A more general concept might be also useful.

\textit{Definition}. For the quadratic, periodic Hamiltonians, even if the evolution in $[0,T]$ does not close to a loop, one might apply (\ref{X}) defining the \textit{Floquet point} $\vect{X}(\tau)=\frac{1}{T×}\int_\tau^{\tau+T}\vect{x}(t)dt$ (though, in general, $\vect{X}(\tau)$ will depend on $\tau$). 

We shall show now, that the behavior of the loop affected by an additional force described in Secs. \ref{sec2}-\ref{sec3}  is a typical phenomenon for all loop processes generated by any quadratic Hamiltonians. 

\begin{proposition}
 Suppose, a time-periodic Hamiltonian $H(t)\equiv H(t+T)$, qua\-dra\-tic in the canonical ob\-ser\-va\-bles $\vect{q}$, generates a loop (\ref{UT0}) in its periodicity intervals $[nT, (n+1)T]$, $n=0, 1, ...$. Then the precession or stability of the loop under an additional, constant force $F$ depends on the loop center $\vect{X}=(X_1,\cdots, X_n)$. If its coordinates vanish, then the loop can change its shape but remains closed and stable. If the coordinates commute, then the center is stable even if the trajectory is not. However, if $\vect{X}$ is a ``fuzzy point'' with non-commuting coordinates, $[X_k,X_l]=i\kappa_{kl}\not\equiv 0$, then the loop will show a drift with a constant velocity $\sim v_k=\kappa_{kl}F_l$ (summation convention) in the direction orthogonal to $F$.
\end{proposition}

\textit{Proof.}  Suppose the perturbing force is $F=(F_1, ...,F_s)$. The perturbed Hamiltonians then read:
\begin{eqnarray}
 \tilde{H}(t)=H(t)-F\vect{x},
\end{eqnarray}
yielding the modified evolution operator $\widetilde{U}(t,0)$ in the customary form:
\begin{eqnarray}
 \widetilde{U}(t,0)=U(t,0)W(t)
\end{eqnarray}
in which $U(t,0)$  is the evolution operator of the unperturbed loop (\ref{UT0}), while $W(t)$  obeys the evolution equation in the interaction frame:
\begin{eqnarray}\label{dW}
 \frac{dW}{dt}=iF\vect{x}(t)W(t); \qquad\qquad W(0)=\mathbb{1}
\end{eqnarray}
where
\begin{eqnarray}
 \vect{x}(t)=U(t,0)^\dagger \vect{x}U(t,0)
\end{eqnarray}
are the time-dependent Heisenberg's observables defined by the unspoiled loop evolution (\ref{BogA20}), (\ref{UT0}). Since the commutators $[x_k(t),x_l(t')]$ for any $t, t'\in\mathbb{R}$ are numbers, the operators $W(t)$ are given by the simplest case of the BCH formulae \cite{Magnus:1954},  \cite{BMP,Bogdan:1970} in which only the integral of $\vect{x}(t)$ matters:
\begin{eqnarray}\label{W}
 W(t)=e^{i\chi(t)}e^{iF\int_0^t\vect{x}(t')dt'},
\end{eqnarray}
and $\chi(t)$ is a real, c-number phase for each $t\in\mathbb{R}$. So,
\begin{eqnarray}\label{40}
 \widetilde{U}(T,0)=U(T,0)W(T)=e^{i\chi}e^{iTF\vect{X}},\qquad \chi\in\mathbb{R},
\end{eqnarray}
implying:
\begin{eqnarray}
 \widetilde{U}(T,0)^\dagger \,F\vect{X}\,\widetilde{U}(T,0)=e^{-iTF\vect{X}}F\vect{X} e^{iTF\vect{X}}=F\vect{X},
\end{eqnarray}
but simultaneously
\begin{eqnarray}
\fl \widetilde{U}(T,0)^\dagger Y\widetilde{U}(T,0)=e^{-iTF\vect{X}}Y e^{iTF\vect{X}}=Y-iT\;[F\vect{X},Y]=Y+\mathfrak{v} T,
\end{eqnarray}
for any $Y=n_lX_l$, where $\mathfrak{v}=n_lx_{lj}F_j$ (summation convention). Hence, the (constant) external force $F$ cannot change the center coordinate $F\vect{X}$, but if $[F\vect{X},Y]\neq 0$ it can shift the combination $Y$ of the remaining ones. If the loop is $n$-times affected by the same $F$, then $F\vect{X}$ is still unchanged, but $Y$ performs a cumulative drift:
\begin{eqnarray}\label{proof}
\fl \widetilde{U}(nT,0)^\dagger Y\widetilde{U}(nT,0)=W(T)^{\dagger n}YW(T)^n=e^{-inTF\vect{X}}Y e^{inTF\vect{X}}=Y+nT\mathfrak{v}
\end{eqnarray}
with the same constant velocity $\mathfrak{v}$.$\fullsquare$\\

\textit{Note}. Above, it is \textit{not assumed} that $T$ is the \textit{smallest} period of $H(t)$, but only that it is a common period of $H(t)$ and of the loop phenomenon. In fact, a typical situation is that the loop occurs after several periods of $H(t)$ (compare Sec. \ref{sec7}).\\

 Our results permit to classify the loops according to the properties of $\vect{X}$:

\begin{enumerate}
 \item[$\bullet$]A loop generated by a quadratic Hamiltonian for $s\geq2$ can have a `fuzzy center' $\vect{X}$ with non-trivial $[X_k,X_l]=i\kappa_{kl}\neq 0$. Then under the influence of a constant external force $F$ the loop can be broken; the former motion center will drift with a constant velocity in a direction orthogonal to the applied force.
 \item[$\bullet\bullet$]If $[X_k,X_l]=0$, then the loop center $\vect{X}$, no matter the external force, must return to its initial value. Various subcases are still possible.
	\begin{enumerate}
	\item[\textit{i}] If $\vect{X}$ does not vanish identically, the loop still can be broken. The motion center won't drift but the trajectory can precess around the fixed center $\vect{X}$. The detailed classification of the loops in this subcase is still an open problem.
	\item[\textit{ii}] If $\vect{X}$ identically vanishes, then $W(T)$ in (\ref{40}) is just a phase factor and the loop is \textit{stable}: it can be deformed but it won't be broken, neither will it precess under the influence of $F$. 
	\end{enumerate}
\end{enumerate}
The sense of this classification turns obvious if one compares the traditional harmonic oscillator with the ``magnetic oscillator'' $H_\beta\; (\beta\equiv $constant). Both admit circular orbits, however, the general elliptic orbit of the oscillator affected by an external force $F$ gets simply displaced in the direction of $F$ (where it remains stationary), while the orbit of $H_\beta$ starts drifting in the direction orthogonal to  $F$. The key to this difference is the distinct nature of the classical/quantum center $(X,Y)$ of both motions. While for the $2D$ oscillator the center $\vect{X}$ is exactly zero, the same center for $H_\beta$ is a fuzzy point (\ref{XyY-X})-(\ref{XyY-Y}) implying the drifting trajectory.

 In physical terms, the difference between both cases has some historical key. In fact, by reading (\ref{hamiltoniano}) inversely: $H_\mathrm{osc} =H_\beta+\frac{\beta}{m}M_z$, one obtains a modern equivalent of an old idea: the description of the elliptic orbit of $H_\mathrm{osc}$ as the superposition of two circular motions, for $H_\beta$ and $M_z$, $i.e.$ the Ptolemean picture of the oscillator trajectory. So, in a sense, ``the epicycles are more stable than cycles''.


\section{Landau fields: manipulating the fuzzy center}\label{sec6}
\begin{figure}[h]
  \centering\includegraphics[width=.65\textwidth]{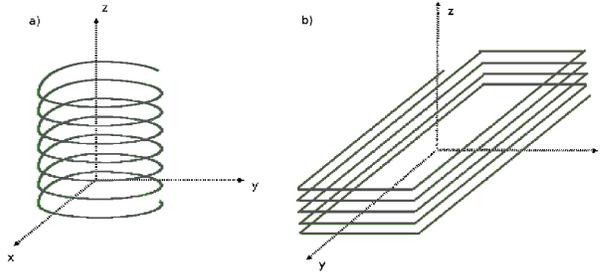}\label{equipotenciales}
\caption{\small a) Cylindrical geometry. b) Landau geometry. \label{fig:geometrias}} 
\end{figure}

As already known, the magnetic operations can produce special effects such as the rigid displacement, squeezing, and the distorted free evolution \cite{Bogdan:1994,Dodonov:2002,Torres:2006,Dodonov:1994}. Below, we shall be specially interested in the homogeneous magnetic fields in a fixed direction (e.g. of $z$-axis), imitating the variable oscillator potentials in 2D \cite{Fish2,Dodonov:1994,Bogdan:1986,Bogdan:1994,Francisco:1999}.
The simplest physical conditions to approximate such fields arise in space domains surrounded by  time dependent currents, e.g., by solenoids of various forms (Fig. \ref{fig:geometrias}a,b). 

If the surface currents do not depend on $z$, the motion in the $x,y$-plane decouples leading to some typical vector potentials in $2D$, e.g.:
\begin{eqnarray}
\fl \quad cylindric:\quad\vect{A}=\frac{B}{2}\left\|\begin{array}{c}\label{cilland}
              -y\\x
             \end{array}\right\| \qquad\qquad Landau:\quad\vect{A}=B\left\|\begin{array}{c}
              -y\\0
             \end{array}\right\|
\end{eqnarray}
If $B$ is static, both expressions (\ref{cilland}) are exact and gauge equivalent. Yet, for variable $B=B(t)$ they still offer a good laboratory approximation, generally used in the ion trap description, though neglecting the retarded field effects in not too huge operation areas\footnote{Differently than for the static $B$, the cylindrical and Landau potentials in (\ref{cilland}) for variable $B(t)$ are no longer gauge equivalent, since they produce different electric fields, corresponding to distinct  geometries of the distant sources; c.f. Dodonov et al. \cite{Dodonov:1994}.}. Below, we shall check the loop behaviour for the time dependent quadratic Hamiltonian of the Landau's case:
\begin{eqnarray}
H(t) &= \label{hamiltoniano-landau} \frac{1}{2m}\left[ \left( p_x+\frac{eB(t)}{c}y\right) ^2+p_y^2\right].
\end{eqnarray}
In the dimensionless variables with $t\rightarrow t\,${\scriptsize T} and $\beta=\frac{eB(t) \mathrm{\scriptsize T}}{2mc}$ (where {\scriptsize T} represent a time unit), it reads:
\begin{eqnarray}\label{hamiltonianolandau}
 H(t)=\frac{1}{2}\left[ \left( p_x+2\beta(t)y\right) ^2+p_y^2\right] 
\end{eqnarray}
The canonical equations
\numparts\label{ecsmovlandau}
\begin{eqnarray}
\frac{dx}{dt} = (p_x+2\beta y) \qquad & \qquad
\frac{dy}{dt} = \;\;\,p_y\label{ecsmovlandaua}\\
\frac{dp_x}{dt} = 0 \qquad & \qquad
\frac{dp_y}{dt}  = -2\beta(p_x+2\beta y).\label{ecsmovlandaub}
\end{eqnarray}\endnumparts
despite their apparent simplicity, require a computer study.  We thus opted to approximate an arbitrary $\beta(t)$ by a step function of $2n$ constant fields $\beta_1, \beta_2, \cdots, \beta_{2n}$ in the subsequent $2n$ time intervals $[t_{i-1},t_i]$, $i=1,\cdots, 2n$ ($0=t_0<\cdots<t_{2n}=T$).  While such steps are never exact (to switch on or switch off an  electromagnetic field takes at least 100 attosecond, \cite{pnu:823},  \cite{pnu:844,Sansone:2006}), our aim here is just to check the principal part of the behaviour. To simplify still further, we fixed the $\beta_i$ and $\Delta t_i=t_{i}-t_{i-1}$ to convert each step into a `$\pi$-pulse' generating a certain semi-circle in the $x, y$ plane, with $\beta_i \Delta t_i=\pm\frac{\pi}{2}$ ($i=1,2,\cdots$). The semicircular fragments of the canonical trajectory correspond to the evolution matrices  
\begin{figure}[b]
\begin{minipage}{0.45 \textwidth} 
\includegraphics[scale=.26, angle=270]{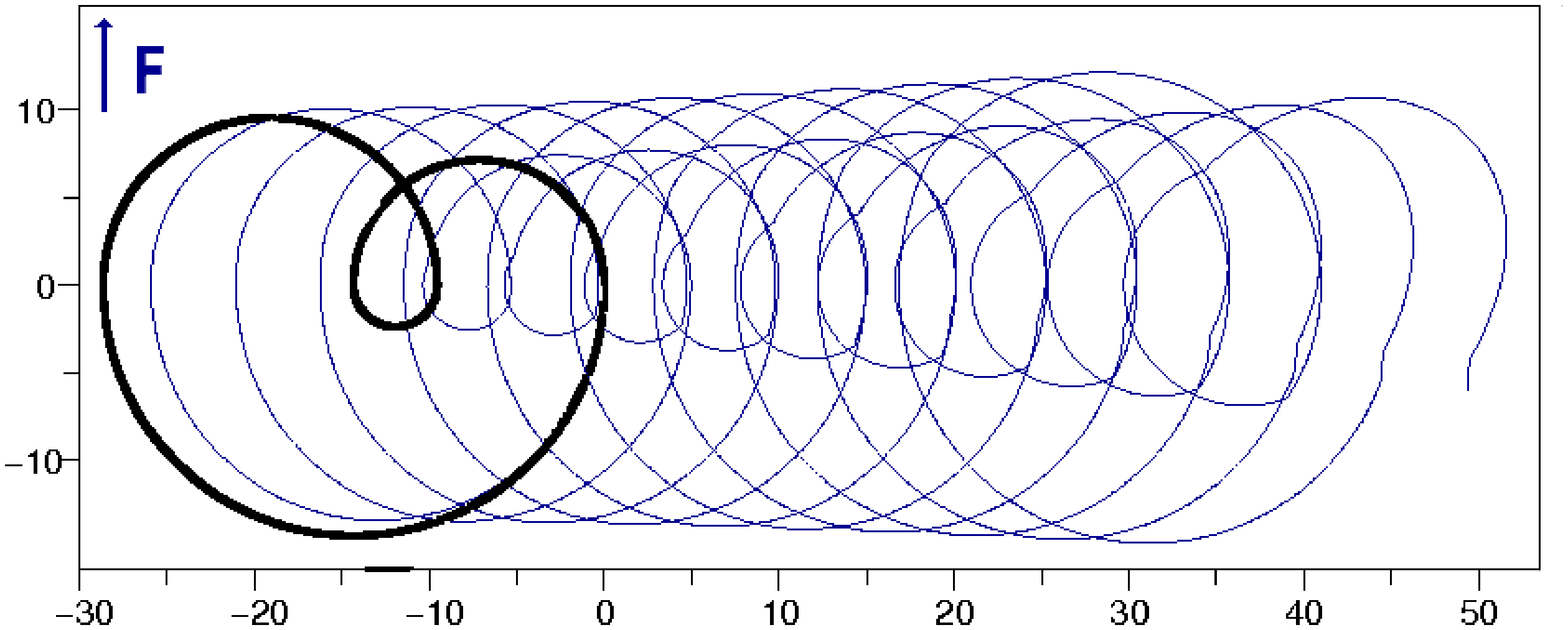}
\end{minipage}
\hfill \begin{minipage}{0.45 \textwidth}%
\begin{center}
\includegraphics[scale=.2, angle=270]{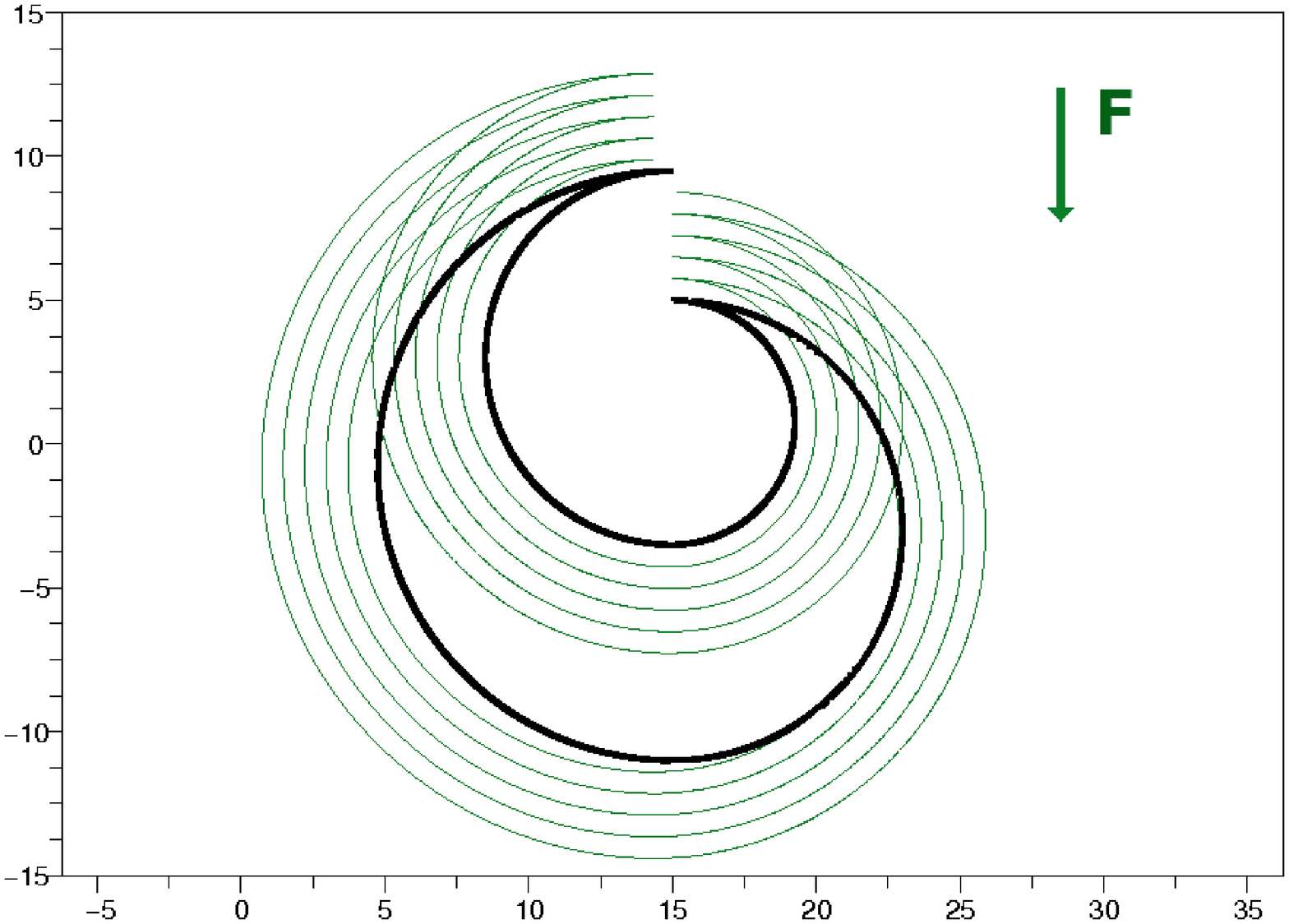}
\end{center}
 \rput(-6.5,4){ {\small $a)$}}
  \rput(.7,4.1){ $b)$}
\end{minipage}
\begin{quote}
\caption{\label{fig:gota1}\small $\vect{a)}$ In black: the classical trajectory illustrating the loop (\ref{ub1b2b3b4}) with  $\beta_1=\frac{\pi}{6}$, $\Delta t_1=3$;  $\beta_2=\frac{\pi}{4}$, $\Delta t_2=2$; $\beta_3=\pi$, $\Delta t_3=\frac{1}{2}$; $\beta_4=\frac{\pi}{3}$, $\Delta t_4=\frac{3}{2}$. The ``fuzzy center'' is $(X,Y)={\left( x + \frac{41}{14\pi} p_y, -\frac{31}{14\pi} p_x + \frac{6}{7\pi^2} p_y\right).} $ In blue: the force $F=(0,1)$ causes the drift of the loop center. $\vect{b)}$ The loop for  $\beta_1=\frac{1}{4}$, $\Delta t_1=2\pi$;  $\beta_2=1$, $\Delta t_2=\frac{\pi}{2}$; $\beta_3=-\frac{1}{4}$, $\Delta t_3=2\pi$; $\beta_4=-1$, $\Delta t_4=\frac{\pi}{2}$. The motion center $(X,Y)=\left( x+\frac{3}{2}p_y, \frac{3}{\pi}p_y\right)$ is operational but no fuzzy. Green: affected by the force $F=(0,-\frac{1}{20})$.}
\end{quote}
\end{figure}
\begin{eqnarray}\label{eulandau}
u_i=u(t_i,t_{i-1})=\left\|\begin{array}{cccc}
\;1 \;& \;  0 \;&\; 0    	\;& \;\frac{1}{\beta_i} \;\\
0 &   -1 & -\frac{1}{\beta_i} 	& 0 \\
0 &		0 &		1 		  	&		0		\\
0 &\;0\;& 	\;0\;		  &  	\;-1\;
    \end{array}\right\|,
\end{eqnarray} 
leading to the final result of $2n$  operations:
\begin{eqnarray}
\fl {u}(T)=u_{2n}\cdots u_1=\mathbb{1}+\Gamma    \label{ub1b2b3b4}
\left\|\footnotesize\begin{array}{cccc}
     &     &     &\;1\; \\
     &     &\;1\;&      \\
     &\;0\;&     &      \\
\;0\;&     &     & 		
 \end{array}\right\|;\qquad  \Gamma=\sum\limits_{i=1}^{2n-1}\left(\frac{1}{\beta_i}-\frac{1}{\beta_{i+1}}\right)
\end{eqnarray} 
with the loop condition ${u}(T)=\mathbb{1}\Leftrightarrow\Gamma=0$.


The simple form of the semicircular operations makes possible not only to draw the trajectories but also to calculate exactly their centers $\vect{X}=(X,Y)$ and commutators $[X,Y]$. As an elementary example, we did it for the 4 step loops obtaining:
\begin{eqnarray}\label{conmutadorlandau}
[X,Y]=-\frac{i\pi}{4T}\left[ \frac{1}{\beta_1|\beta_1|}+\frac{1}{\beta_2|\beta_2|} +\frac{1}{\beta_{3}|\beta_{3}|} +\frac{1}{\beta_{4}|\beta_{4}|}\right] .
\end{eqnarray}
Within the loop condition $\Gamma=0$, the commutator  (\ref{conmutadorlandau}) admits still various control options, producing the ``fuzzy centers'' with distinct drifting capacities (see Fig~\ref{fig:gota1}a). However,  $[X,Y]$ may also vanish for some values of $\beta_i$. This happens $e.g.$ for $\beta_3=-\beta_1$, $\beta_4=-\beta_2$  when the quantum center is operational, but not fuzzy (see Fig.~\ref{fig:gota1}b), the case which cannot occur for static fields.

From the relativistic point of view the Hamiltonians (\ref{hamiltoniano-landau}) are inexact, since the retarded effects are missing. Notice, however, that the non-relativistic Landau's potential (\ref{cilland}) with $B=B(t)$ is the limit for $\frac{1}{c}\rightarrow 0$ of an exact, relativistic expression in form of a finite difference:
\begin{eqnarray}
\fl \vect{A}(t,\vect{x})=\left\|
\begin{array}{c}
A(t,\vect{x})\\
0
\end{array}\right\|; \qquad A(t,\vect{x})=\frac{1}{2}\frac{G(t-\frac{y}{c})-G(t+\frac{y}{c})}{\frac{1}{c}} \;\,\rightarrow-G'(t)y
\end{eqnarray} 
of two plane \textit{Landau pulses} propagating in opposite directions, where $G(t)$ is a continuous, real function with a bounded, piece-wise continuous derivative $G'(t)=B(t)$, modeling the time dependent magnetic fields for $\frac{1}{c}\rightarrow 0$. So, the non-relativistic Hamiltonian (\ref{hamiltoniano-landau}) with the time-dependent $B(t)$ is indeed the $1^{st}$ step of the Einstein-Infeld-Hoffman (EIH) approximation commonly used in General Relativity to describe the slow motions in limited space domains \cite{EIH}. In the laboratory of the size of $\approx 1m$ the  $y$-dependent delays don't exceed $\delta t=\frac{1m}{c}\cong \frac{1}{3}\cdot 10^{-8}sec$ and the approximation is almost perfect (errors invisible in our Fig. \ref{fig:gota1}a,b). 
We thus conclude that in not too huge magnetic traps the ``fuzzy'' centers of the Hamiltonian (\ref{hamiltoniano-landau}) describe correctly the principal part of the drifting phenomenon.

\section{Cylindrical geometry: the resistant loops }\label{sec7}

In turn, we shall examine some curious phenomena induced by softly pulsating fields  of cylindrical geometry. The non-relativistic Hamiltonian becomes:
\begin{eqnarray}
H (t) =\label{hamiltoniano-cilindrico} \frac{1}{2m}\left[\vect{p}^2+\left(\frac{eB(t)}{2c}\right)^2\vect{x}^2\right]-\left(\frac{eB(t)}{2mc}\right)M_z.
\end{eqnarray}
 Differently than the `homogeneous Schr\"odinger's equation' (c.f. \cite{Belli}), the eq. (\ref{hamiltoniano-cilindrico}) represents a $2D$ `Aristotelian world', whose symmetry center $x=y=0$ is distinguished by the circular electric fields (c.f. also \cite{Dodonov:1994}).  In principle, $B(t)$ might be arbitrary, but below, we shall be most interested in the \textit{harmonic} and \textit{biharmonic} fields:
\begin{eqnarray}
 B(t) &= B_0 + B_1\sin(\omega t), \label{uniarmonicos} \\[5pt]
 B(t) &= B_1\sin(\omega t) + B_2\sin(2\omega t), \label{biarmonicos}
\end{eqnarray}
of period $T=\frac{2\pi}{\omega}$.  Following the commonly applied approximation \cite{Malkin:1970,Paul:1993,Dodonov:2002} we adopt the semiclassical picture (Thomson rather than Compton \cite{Resnick:1988}). Our design is quite unsophisticated comparing with the time dependent mass \cite{Caldirola:1941,Kanai:1948} or the variable material constants \cite{DKN,Castanos:1996}. However, what precisely happens with the microparticles in this simple scenario? 

 To describe the ample classes of similar evolution processes it is practical to introduce the dimensionless time, 
$t\rightarrow\frac{\omega}{2\pi}t=\frac{t}{T}$, field $ \beta=\frac{\pi e}{m \omega c}B$, and canonical variables $[x,p_x]=[y, p_y]=i$. The field oscillations now become:
\begin{eqnarray}
 \beta(t)=\beta_0+\beta_1 \sin(2\pi t)+\beta_2 \sin(4\pi t),\label{betacompleto}
\end{eqnarray}
 where $\beta_2=0$ corresponds to the harmonic and $\beta_0=0$ to the biharmonic cases (\ref{uniarmonicos}) and (\ref{biarmonicos}) respectively. The rescaled Hamiltonian is
\begin{eqnarray}\label{HamiltonianoMQ2}\index{Hamiltoniano!cu\'antico}
 H(t) = \underset{H_\mathrm{osc}}{\underbrace{\frac{1}{2}\Bigl(\vect{p}^2+\beta(t)^2\vect{x}^2\Bigr)}}- \underset{H_\mathrm{rot}}{\underbrace{\beta(t) M_z\vphantom{\frac{1}{2}}}}\,,
\end{eqnarray}
with $H_\mathrm{osc}$ representing the time dependent ``magnetic oscillator'' and 
$M_z$ the rotation generator. 
 Since $H_\mathrm{osc}(t)$ and $M_z$ commute, $U(t)$ factorizes into two commuting unitary operators,
$
 U(t)=U_\mathrm{osc}(t)\,U_\mathrm{rot}(t),
$
where:
\begin{eqnarray}\label{ecforU0}
 \frac{dU_\mathrm{osc}(t)}{dt} = -\frac{i}{\hbar} H_\mathrm{osc}(t) U_\mathrm{osc}(t); \qquad\qquad U_\mathrm{osc}(0) =\mathbb{1},
\end{eqnarray}
while
\begin{eqnarray}\label{R}
 U_\mathrm{rot}(t) = e^{-i\int_0^t \beta(t')dt'M_z}
\end{eqnarray}
produces just the rotations $r(t)$ between the canonical pairs $x,p_x$ and $y, p_y$. The canonical transformation defining the evolution matrix $u(t)$ can be split into two steps:
$
\vect{q} \rightarrow  \vect{q}_\mathrm{osc}(t) \rightarrow \vect{q}(t),
$
implemented by $U_{ osc}(t)$ and $U_\mathrm{rot}(t)$ respectively. The operation $U_\mathrm{osc}(t)$ is reducible, affecting separately both canonical pairs $x,p_x$ and $y,p_y$. which evolve simultaneously according to the same $2\times 2$ matrix further denoted $b(t)$, i.e.:
    \begin{eqnarray}
    U_\mathrm{osc}(t)^\dagger\left\|\begin{array}{c}
                 x \\ p_x
                \end{array}\right\| U_\mathrm{osc}(t) =b(t)\left\|\begin{array}{c}
                 x \\ p_x
                \end{array}\right\|\label{a},
    \end{eqnarray}
\textit{idem} for $y, p_y$. 
 By differentiating both sides of (\ref{a}) in agreement with (\ref{ecforU0}) and using the canonical commutation rules, one sees that $b(t)$ is determined by the differential matrix equation:
\begin{eqnarray}
 \frac{db}{dt} = \Lambda(t)b(t),\quad\quad\Lambda(t)=\left\|\begin{array}{cc}
                 0 & 1\;\\ -\beta(t)^2 & 0\; \label{eqforb}
                \end{array}\right\|; \quad\quad b(0)=\mathbb{1}
\end{eqnarray}
 which is at the bottom of all quantum control problems for time dependent oscillator Hamiltonians \cite{Lewis:1967,Lewis:1968,Malkin:1970,Royer:1985,Bogdan:1986,Bogdan:1994,MRPS}. Once having $b(t)$, one immediately constructs $u_\mathrm{osc}(t)$, as the simple pair of two $b(t)$-cells. In turn, multiplying $u_\mathrm{osc}(t)$ by the $4\times 4$ matrix $r(t)$ rotating by $\gamma(t)=\int_0^t\beta(t')dt'$ around the $z$-axis, one obtains the complete $4\times 4$ evolution matrix 
\begin{eqnarray}\label{uxrot}
u(t)= r(t)\;\left\|\begin{array}{cc}
                 b(t) & \\ & b(t)
                \end{array}\right\|.
\end{eqnarray}

Here the role of $b(t)$ is specially relevant for the  periodically repeated field patterns, since the one-period-evolution step $b(T)$ decides about the bounded (stable) character of the motion, or  its capacity to produce the parametric resonance  \cite{Eastham:1974,Magnus:1979,Reed:1975}.

In our case (\ref{betacompleto}), $T=1$. As representing the canonical evolution, $b(t)$ are simplectic, including the Floquet matrix $b(1)$, which permits to classify the motions. Since Det$\big[b(1)\big]= 1$, its eigenvalues depend on just single (real) trace invariant. The characteristic equation
\begin{eqnarray}
 D(\lambda)=\mathrm{Det}\big(\lambda-b(1)\big)=\lambda^2-\lambda\,\mathrm{Tr}\,b(1)+1=0\label{Det}
\end{eqnarray}
has two non-vanishing roots
\begin{eqnarray}
 \lambda_{\pm} = \frac{1}{2}\mathrm{Tr}\,b(1)\pm i\sqrt{\Delta};\qquad
\Delta = 1-\frac{1}{2}\Big[\mathrm{Tr}\,b(1)\Big]^2,\label{roots}
\end{eqnarray}
with $\lambda_{+}\lambda_{-}=1$, distinguishing three possible types of motion:

\renewcommand{\labelenumi}{\Roman{enumi}.}
\begin{enumerate}
 \item \textit{Stability area}. If $\big|\mathrm{Tr}\, b(1) \big|<2$, then $\lambda_{+}, \lambda_{-}$ are two different complex eigenvalues with $|\lambda_{+}|=|\lambda_{-}|=1$ (phase factors) of the form $\lambda_{+}=e^{i\varphi}, \lambda_{-}=e^{-i\varphi}, \varphi\in\mathbb{R}.$ The $b(1)$ shows an osci\-lla\-ting behaviour of 
$
 b(n)=b(1)^n, n\in\mathbb{N}
$
defining the bounded trajectories.\\[-7pt]
\item The \textit{threshold} (\textit{separatrix}) is characterized by $\big|\mathrm{Tr}\,b(1)\big|=2$. The Floquet matrix $b(1)$ here has two coinciding eigenvalues $\lambda_{+}=\lambda_{-}=\pm1$. If $b(1)$ is diagonalizable then once again $b(t)$ trajectories are bounded, but if not, then they can show a weak parametric resonance growing in arithmetic but not geometric progression.\\[-7pt]
\item If $\big|\mathrm{Tr}\,b(1)\big|>2$, then $b(1)$ has a pair of real eigenvalues, $\lambda_{+}, \lambda_{-}\neq 0$, $\lambda_{-}=\frac{1}{\lambda_{+}}$, of which at least one has the absolute value $>1$. The trajectories show a strongly resonant behavior of the squeezing type. 
\end{enumerate}
\renewcommand{\labelenumi}{\arabic{enumi}.}

In the best known case of ion traps with the sinusoidally oscillating elastic force  the solutions $b(t)$ are  expressed in terms of the \textit{Mathieu functions} and the resonance borders form the well known Strutt diagramme (see e.g.\cite{Bender:1978,MRPS}). Due to its familiar shape, most of studies illustrating the effects of the time dependent elastic potentials stick to the Mathieu scheme (see $e.g.$ Paul's trap \cite{Paul:1993}, the quantum tomography \cite{Manko:1997}, etc.). The need of some wider stability designs was pointed out by Glauber (c.f. the statement in Baseia et al \cite{Ba}). Indeed, some distinct stability cases were considered for finite dimensional state spaces \cite{Pogo1,Castanos:2005,Castanos:2006}, or else, for the rotating magnetic \cite{David:1990,BD89,David:1992c} or electric fields \cite{IBB,Sara:2005}. However, this does not include the magnetic oscillators $H_{osc}$ (\ref{HamiltonianoMQ2}) modulated by $\beta(t)^2$. The difference seems modest but the consequences are not.

\subsection{Harmonic loops: an anomalous resistance?}
\begin{figure}[t]
\centering\includegraphics[width=\textwidth]{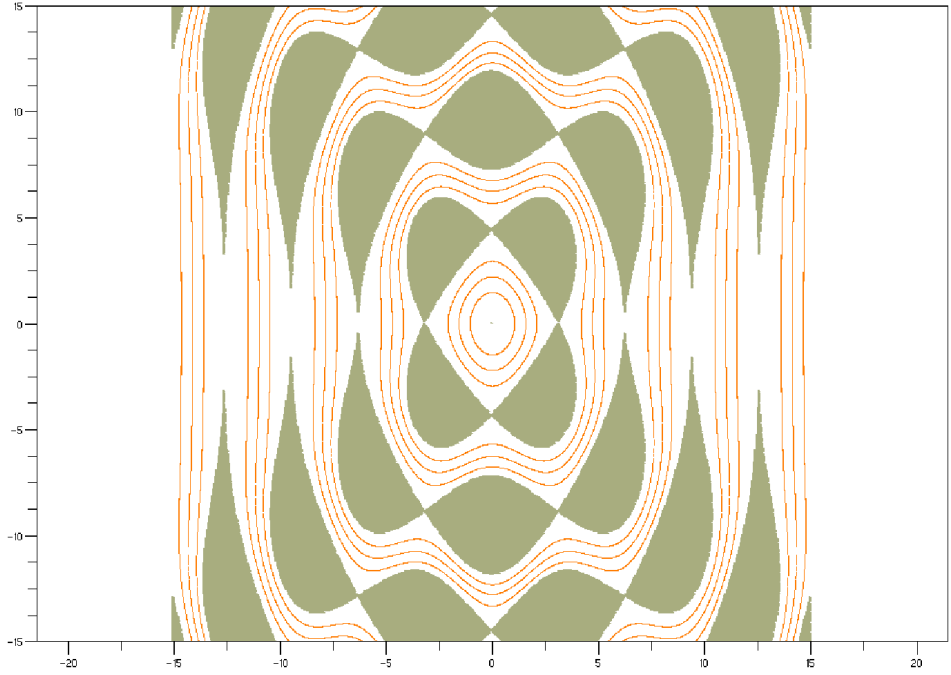}
\begin{quote}
 \caption{\label{fig:delgadomap}\small The types of Floquet operations $U(1)$ generated by the  harmonic  case of (\ref{betacompleto}) on the $(\beta_0, \beta_1)$ plane (the map of Delgado). The clear areas are the stability zones, in which the \textit{evolution loops} can occur for $\mathrm{Tr}\,b(1)=\pm2\cos{\frac{2\pi l}{n}}$, under the subsidiary condition $r(1)=1$.}
\end{quote}			
\end{figure}

  A  systematic study of the harmonic magnetic case (\ref{uniarmonicos}) was undertaken by F. Delgado \cite{Francisco:1999,Bogdan:1998} by scanning the stability thresholds for the time dependent magnetic oscillator (\ref{HamiltonianoMQ2}) with the harmonic $\beta(t)$. The result was one of the first maps on the plane of the dimensionless amplitudes $(\beta_0,\beta_1)$, which differs notably from the Strutt diagramme (see Fig. \ref{fig:delgadomap}). 

\begin{figure}[t]%
 \parbox{1in}{\hspace*{-.8cm}\includegraphics[scale=.3, angle=270]{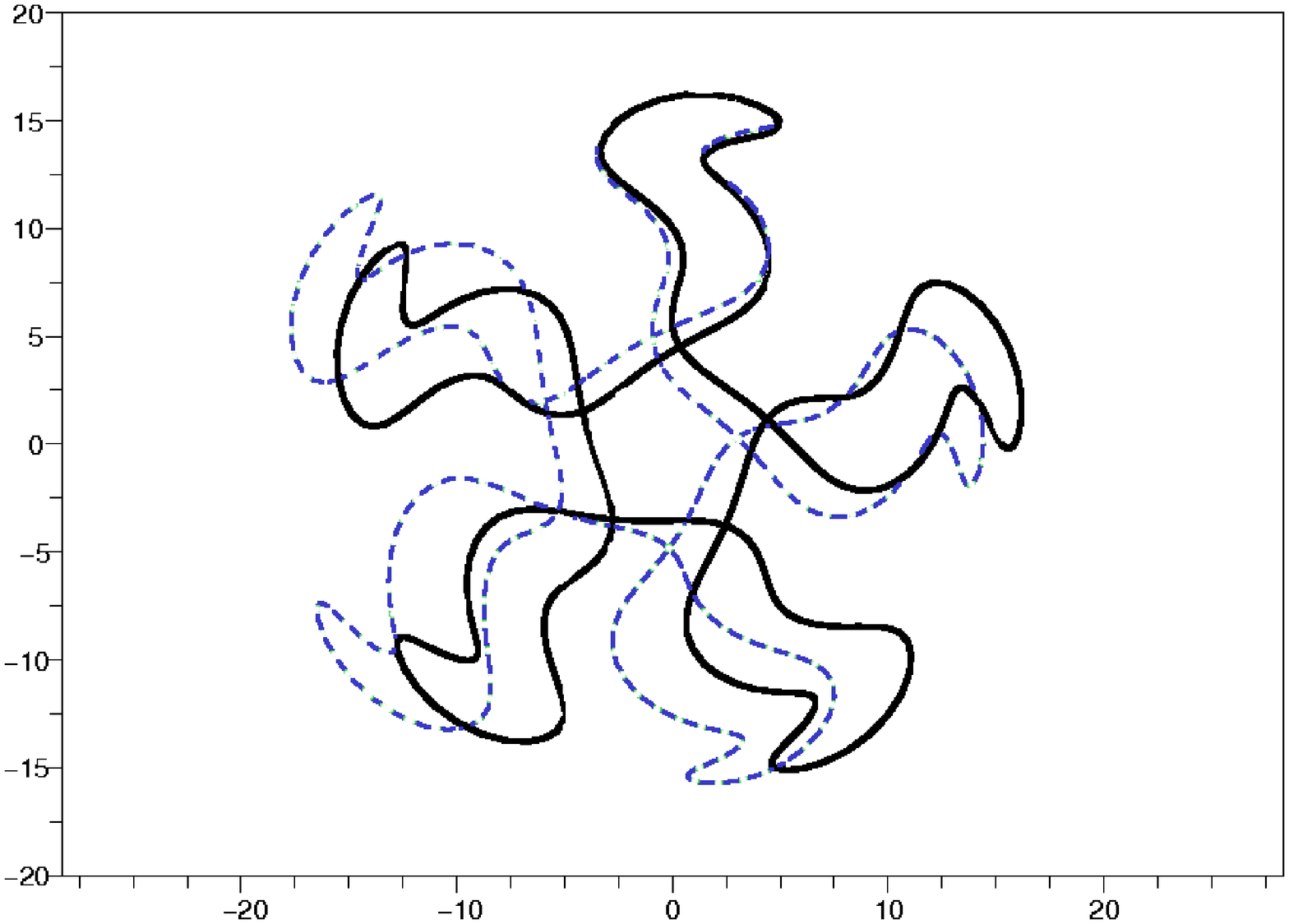}}%
 \qquad\qquad\qquad\qquad\quad\quad
 \begin{minipage}{1in}%
 \includegraphics[scale=.3, angle=270]{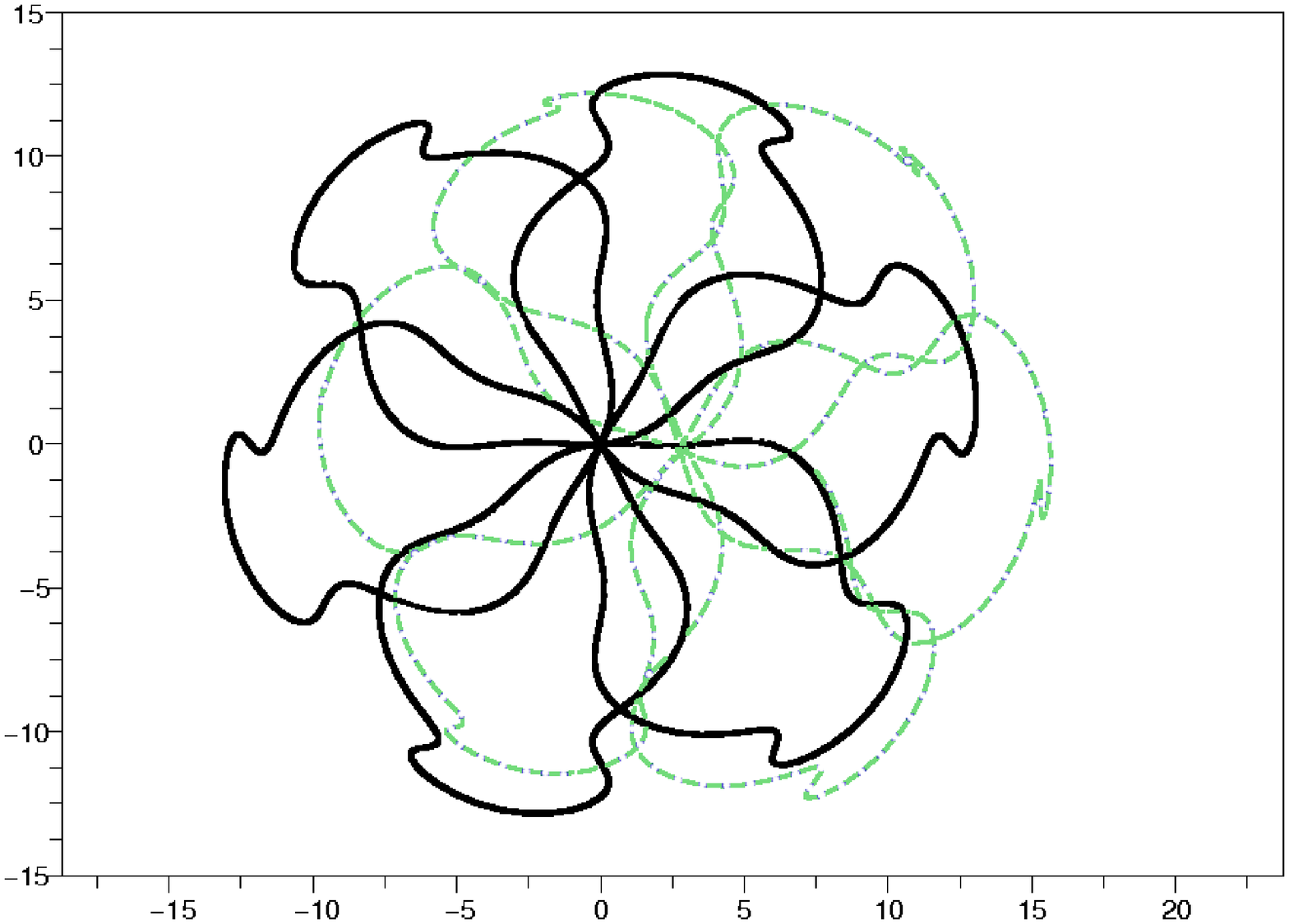}
   \rput(-6.4,5.75){ $a)$}
  \rput(1.3,5.75){ $b)$}
 \psline[linewidth=0.3mm,linecolor=Verde]{->}(6,4.75)(6.7,4.75)
 \rput[cm](6.4,4.5){\textcolor{Verde}{$\vect{F}$}}
 \psline[linewidth=0.3mm,linecolor=blue]{->}(-5,4.75)(-5.7,4.75)
 \rput[cm](-5.4,4.5){\textcolor{blue}{$\vect{F}$}}
 \end{minipage}%
 \begin{quote}
  \caption{\label{fig:evolutionloop15}\small The classical trajectories in the $xy$ plane illustrating the  evolution loops. $a)$ For $\beta(t)=-\frac{\pi}{5}-1.152\sin{(2\pi t)}$, closing after 15 periods. In blue the deformation (but not drifting) under the constant force $\vect{F}={(-1.5,0)}$. $b)$ The evolution loop for $\beta(t)=\frac{\pi}{8}-0.815\sin{(2\pi t)}$ closing after 24 field periods. In green, the deformation by the constant force $\vect{F}={(1,0)}$.}
 \end{quote}
 \end{figure}

By looking for $\mathrm{Tr}\,b(1)=\pm2\cos{\frac{2\pi l}{n}}$, the computer identified also a sequence of curves in the stability areas where
\begin{eqnarray}
  \lambda_{\pm}=e^{\pm\frac{2\pi i\,l}{n}}
\end{eqnarray}
 granting that oscillatory motions close after the $n$ periods of $H(t)$, forming the twin loops in both $2D$ subspaces $(x, p_x)$ and $(y, p_y)$. In order to generate the loop of 4 canonical variables, one must assure that the rotation $r(1)$ simultaneously closes. The harmonic component $\beta_1 \sin(2\pi t)$ does not contribute to $r(1)$. The only condition is that the constant intensity $\beta_0$ should rotate the canonical variables by $\pm2k\pi$ after some $m$ repetitions ($k, m=1,2,...$). This distinguishes the sequence of straight lines $\beta_0=\pm\frac{2\pi k}{m}$ on the stability map of Fig \ref{fig:delgadomap}. Their intersections with the loop curves yield the amplitude pairs $(\beta_0, \beta_1)$ generating the loop phenomena for all 4 canonical variables after a finite number of $m n$ repetitions (see Fig.~\ref{fig:evolutionloop15}).

While the existence of the harmonic loops is known \cite{Francisco:1999}, their extremely regular, kaleidoscopic forms have some more implications. One of them is the exact vanishing of the ``operator centers'' $\vect{X}=(X,Y)$. Indeed:

\begin{proposition}
 Suppose, for a pair of amplitudes $(\beta_0, \beta_1)$ the evolution loop of the oscillatory part $b(t)$ closes for $\tau=n$, while the rotation $r(\tau)$ yields a certain non-trivial angle $\theta=\left( \frac{k}{m}\right)2\pi$, where $\frac{k}{m}$ is rational but not an integer (no matter, whether $\frac{k}{m}$ is smaller or greater than $1$). Then, the loop obtained for $t=m\tau=nm$ by simultaneous closing of both, oscillatory and rotational motions has the trivial center on both classical and quantum levels.
\end{proposition}

\textit{Proof.} In fact, the rotation $r(\tau)$ breaks the oscillatory loop at $t=\tau$, marking a new end point $\vect{q}_\theta$  rotated by $\theta$ with respect to the initial $\vect{q}$. The $\vect{q}_\theta$, in turn, becomes the initial point for the next fragment of the trajectory, which again does not close, but is just the $\theta$-rotated version of the previous one, ending up at $\vect{q}_{2\theta}$, etc. As the result, the whole trajectory is the sum of similar fragments, generated between the time moments $0, \tau, 2\tau, ...$ joining the subsequent points $\vect{q}, \vect{q}_\theta, \vect{q}_{2\theta}, \cdots$, each one just the $\theta$-rotated version of the previous one, until finally, at $t=m\tau$,  the trajectory closes with $\vect{q}_{m\theta}=\vect{q}$.
As the sum of the $\theta$-rotated steps, the whole loop is invariant under the $\theta$-rotation and so is the loop center $\vect{X}$ defined by (\ref{Apromedio}). However, the only vector (with numerical or operator components) invariant under a non-trivial rotation around the coordinate center is $\vect{X}=0$.$\fullsquare$\\[10pt]

Consistently with Sec. \ref{sec5} this assures the stability of the harmonic loops which resist drifting (see  Fig. \ref{fig:evolutionloop15}). It looks almost as an elementary analogue of the anomalous resistance  observed in $2D$ \cite{Zudov:2006} (though the attempts of overestimating the analogy might be misleading).

\subsection{The biharmonic case}
In turn, we shall consider the biharmonic fields (\ref{biarmonicos}) or (\ref{betacompleto}) with $\beta_0=0$, defined by pairs of the dimensionless amplitudes $\beta_1, \beta_2$.  Their agreeable property is that at each $t=nT=n$ all rotations cancel, and simultaneously, the field amplitudes $\beta(n)\equiv 0$. Henceforth, the evolution operators $U(nT)=U(n)$ are reduced to the purely oscillatory part; the corresponding one-period evolution matrix (\ref{a}) 
reduces to a pair of $b(1)$ cells. Moreover, the physical sense of the corresponding evolution in the time moments $nT=n$ $(n\in\mathbb{Z})$ is not affected by the difference between the canonical and kinetic momenta. 

The stability areas determined by the computer scanning of $\mathrm{Tr}b(1)$ (c.f. \cite{Adoc}) form now a new 2-parameter map (see Fig.~\ref{fig:bicafeverde}).
\begin{figure}[t]
\centering\includegraphics[height=.7\textwidth,angle=-90]{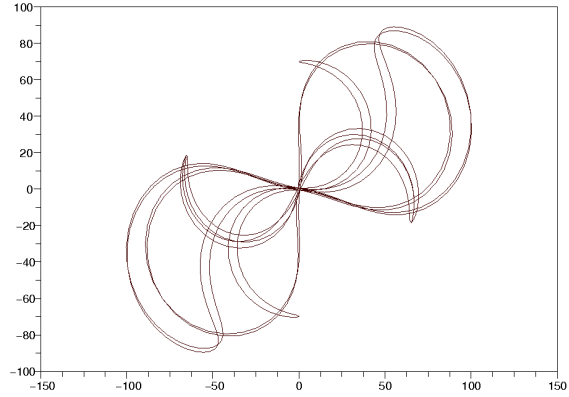}
 \begin{quote}\caption{\label{fig:bi6}\small Biharmonic evolution loop for $\beta(t)=\frac{\pi}{2}\sin(2\pi t)+9$.$966\sin(4\pi t)$  on $x,y$-plane. The loop closes  after 6 field periods. }
 \end{quote}
\end{figure}
As before,they are densely populated by curves $\mathrm{Tr}\,b(1)=\pm2\cos{(\frac{2k\pi}{n})}$, whose points $(\beta_1, \beta_2)$ generate loops, though now  $(\beta_1, \beta_2)$ don't need to fulfill any extra condition as the rotations $r(1)$ automatically cancel for $t=n\in\mathbb{Z}$. A loop which closes up after 6 field periods is drawn on Fig.~\ref{fig:bi6}. Note its  symmetry under the parity reflection, which grants the exact vanishing of $\vect{X}$. However the parity is not even necessary.

\begin{proposition}
 Consider a periodic sequence of field pulses $\beta(t)\equiv\beta(t+1)$,  with $\beta(k)=0,$ for $k\in\mathbb{Z}$, and moreover, $\int_0^1\beta(t)dt=0$ ($i.e.$, the rotations cancel in each full period $T=1$ and its multiples). Then the evolution loops which close for the first time after a certain number $n>1$ of the $\beta$-periods has the vanishing center $\vect{X}=0$.
\end{proposition}

\textit{Proof.} Within our assumptions the eigenvalues $\lambda_{\pm}$ of $b(1)$ are the $n$-$th$ roots of unity, $\lambda_{+}^n=\lambda_{-}^{n}=1$. If now $\lambda_{+}\neq\lambda_{-}$, then $b(1)$ is diagonalizable, with complex eigenvalues, and so is the $4\times 4$  matrix $u(1)$. If $\lambda_{+}=\lambda_{-}$ the only possibilities are $\lambda_{+}=\lambda_{-}=1$ and $\lambda_{+}=\lambda_{-}=-1$. Should then $b(1)$ be non-diagonalizable, it 
could not produce the closed process, contrary to our assumption. Hence, the cell must be trivial and $b(1)$ diagonalizable. If so,  $\lambda_{+}=\lambda_{-}=1$ $\Rightarrow$ $b(1)=\mathbb{1}$ is excluded by the assumption that the loop closes only after a number $n>1$ of steps. Hence, either $b(1)=-\mathbb{1}$, but if not, then $b(1)$ is a real matrix with a pair of different complex eigenvalues. The corresponding $u(1)$ has either $4$ identical eigenvalues ($-1$) or two (identical) essentially complex pairs. The evolution matrix $u(t)$ in the sequence of time intervals $[0,1)\cup [1,2) \cup \cdots \cup[n-1,n]$ reproduces the $u(t)$ from $[0,1)$, but preceded by the increasing powers of $u(1)$:\\

 \vspace*{1.9cm}
 \hspace*{.34\textwidth}\hexagonoc{1.5cm}

 \newpage
If now $u(1)^n=\mathbb{1}$, then the whole matrix trajectory is invariant under the multiplication by $u(1)$ and so is the trajectory center $\bm{X}$ which must fulfill $u(1)\bm{X}=1\bm{X}$; but since $u(1)$ has no eigenvalue 1, the center must vanish identically, implying the loop stability (the loop resists the external forces).$\fullsquare$\\

The group theoretical sense of the $2D$-loops can be noticed by reinterpreting the time variable $\theta=2\pi t$ as the rotation angle in the plane. The quantity which returns to its initial value after a full $2\pi$-rotation is a tensor. An entity which changes its sign is a spinor. An entity which returns to its initial value only after a finite number of $2\pi$-rotations (as e.g. the loops of Figs. \ref{fig:evolutionloop15} and \ref{fig:bi6}) obeys a fractional representation of the rotation group in $2D$ \cite{Juan}. 

\subsection{The imperfections}
To estimate the corrections to the scheme it is convenient to return to the ordinary units.
For slow classical/quantum motions (low kinetic momenta), the use of the Schr\"odinger's QM seems justified; but some dissipative correction must affect our semiclassical results. If the pulsating field is exactly periodic, the probability of quantum jumps might be read from the quasienergies of the Floquet Hamiltonian. Yet, for the low frequency of the applied fields and the evolution limited to a few pulse periods, more reliable seems the Abraham-Lorentz radiative acceleration $a_{rad}=\sigma\tdot{q}$, where $\sigma=\frac{2e^2}{3mc^3}$ is the ``characteristic time'' of the charged particle (c.f. Jackson \cite{Jackson}). Taking as the model the electron with $\sigma\cong2.12\times10^{-24}sec$ one has in the dimensionless variables: $a_{rad}=\frac{\sigma}{T}\tdot{q}$; so in the harmonic field frequency corresponding to the long $\approx 1km$ RW the single operation period $T=\frac{2\pi}{\omega}\approx3.3\times10^{-6}$ implying very little A-L corrections $a_{rad}\approx6.4\times10^{-18}\tdot{q}$ (though the evaluation changes for increasing $\omega$; see also \cite{MRPS}).

A different source of errors are the retardation free vector potentials in the laboratory approximations (\ref{cilland}). Their relativistic equivalents can be obtained by looking for the freely propagating fields of cylindrical symmetry:
\begin{eqnarray}
 \vect{A}_\omega(t,r)=\sin\omega t\;\Upphi(r)\left\|\begin{array}{c}
                                          -y\\x
                                       \end{array}\right\|;\qquad
 r=\sqrt{x^2+y^2}.\label{nuevopotencial}
\end{eqnarray}
The d'Alembert eq. $\square\vect{A}_\omega=0$ then traduces into:
\begin{eqnarray}
 \left[\frac{d^2}{dr^2}+\frac{\omega^2}{c^2}\right]r\,\Upphi +\frac{d}{dr}\Upphi\equiv0\label{dalambertODE}
\end{eqnarray}
(compare \cite{DKN93}) which admits an analytic solution $\Upphi(r)=\Upphi_0+\Upphi_2\,r^2+\Upphi_4\,r^4+\cdots$ with the coefficients given by the recurrence $\Upphi_{2n}=-\left(\frac{\omega}{2c}\right)^2\frac{1}{n(n+1)}\Upphi_{2(n-1)}$, $n=1,2,\cdots$. They show that the ``laboratory Hamiltonian'' (\ref{hamiltoniano-cilindrico}-\ref{HamiltonianoMQ2}) is once again the first step of the EIH method \cite{EIH} for the potential (\ref{nuevopotencial}-\ref{dalambertODE}). Its precision depends on the ratio between the effective size $\Delta r$ of each loop and the wavelength associated with the pulse frequency $\omega$, $i.e.$ $\lambda=cT=c\frac{2\pi}{\omega}$. If $\lambda$ is much greater than the loop size, $\frac{\Delta r}{\lambda}\ll1$, then the formula (\ref{nuevopotencial}) describes the harmonic magnetic fields differing little from the homogeneous field (\ref{biarmonicos}), drawing with a good accuracy the regular loop forms predicted by the quadratic Hamiltonian (\ref{hamiltoniano-cilindrico}). Note that an analogous estimation (of the ratio between the trajectory size and the field inhomogeneity parameter) works also in different circumstances, $e.g.$ for the atomic or molecular systems irradiated by the coherent laser light. The typical atomic size ($e.g.$ of widely studied $Rb$ atoms) is in the range $\approx 10^{-8}cm (\sim 1\mathring{A})$ but the $\lambda$ of visible light is much greater (between $3800\mathring{A}$ and $7800\mathring{A}$); which seems one of the reasons why the approximate description of the electric forces by a homogeneous, pulsating field $E\sim E_0\sin{(\omega t)}$ gives so good results in the description of Rabi rotations \cite{Rabi:1954}. The question whether the effects we described can have some analogues for the exact relativistic solution in strongly inhomogeneous fields (e.g. \cite{IBB:2004}) is open.

Until now,  our exploration was limited to the stability areas with their typical ``fauna'' and ``flora'' of drifting and stable loops. However, the problems of quantum control unavoidably leads us across the borders. 

\section{The threshold: $\delta$-kicks and inverted free evolution.}\label{sec8}

Since the rotations commute with the whole Hamiltonian (\ref{HamiltonianoMQ2}), we shall again fix attention on the $b(t)$ cells of the evolution process (\ref{uxrot}).

One of the major control challenges are the idealized $\delta$-kicks of the elastic potentials and the events of the ``distorted free evolution'', both represented by extremely simple operators and $q,p$-transformations represented by $b(1)$:
\begin{eqnarray}
\fl\begin{array}{ccc}
    e^{-ia\frac{q^2}{2}} &\; \rightarrow \;& \left\|\begin{array}{cc}\label{kickdeltas}
              \;\, 1 \;\, & \;\, 0 \;\,\\- a  & 1
             \end{array}\right\|
\end{array},
\quad\begin{array}{ccc}
  \quad e^{-i\tau\frac{p^2}{2}} &\; \rightarrow \;& \left\|\begin{array}{cc}
              \;\, 1 \;\, & \;\, \tau\;\,  \\ 0 & 1
             \end{array}\right\|
\end{array},\quad (\tau\neq T),
\end{eqnarray}
(where $q$, $p$ mean $x, p_x$ or $y, p_y$) though almost impossible to achieve straightforwardly.

The left operation in (\ref{kickdeltas}) can be interpreted as a formal result of an infinitely strong pulse of an oscillator potential applied within an infinitesimal time. The right one, if $\tau>0$, $\tau\neq1$,  represents the distorted (slowed or accelerated) free evolution; but if $\tau<0$, it yields highly counter-intuitive effect: it inverts the free evolution of a wave packet, including the interference between the different packet components. No less provocative are the evolution operators and matrices:
\begin{eqnarray}
\fl\begin{array}{ccc}
    Pe^{-ia\frac{q^2}{2}} &\; \rightarrow \;& -\left\|\begin{array}{cc}\label{ODD}
             \;\, 1 \;\, &  \\ \;\,-a\;\,  & \;\, 1\;\,
             \end{array}\right\|
\end{array},
\quad\begin{array}{ccc}
  \quad Pe^{-i\tau\frac{p^2}{2}} &\; \rightarrow \;& -\left\|\begin{array}{cc}
              \;\, 1 \;\, & \;\, \tau\;\,  \\  & 1\;\,
             \end{array}\right\|
\end{array},
\end{eqnarray}
$i.e.$ (\ref{kickdeltas}) superposed with the parity reflection $P:q\rightarrow -q, p\rightarrow -p$. All matrices (\ref{kickdeltas}), (\ref{ODD}) belong to the resonance borders (case II of our classification). In principle, they can be generated by combinatorial field patterns \cite{Bogdan:1994,Dodonov:1994,DM}. However could they be caused by a ``soft persuasion'' instead of brutal force? 

\begin{sidewaysfigure}
\centering
 \begin{pspicture}(0,15)
\includegraphics[width=20cm]{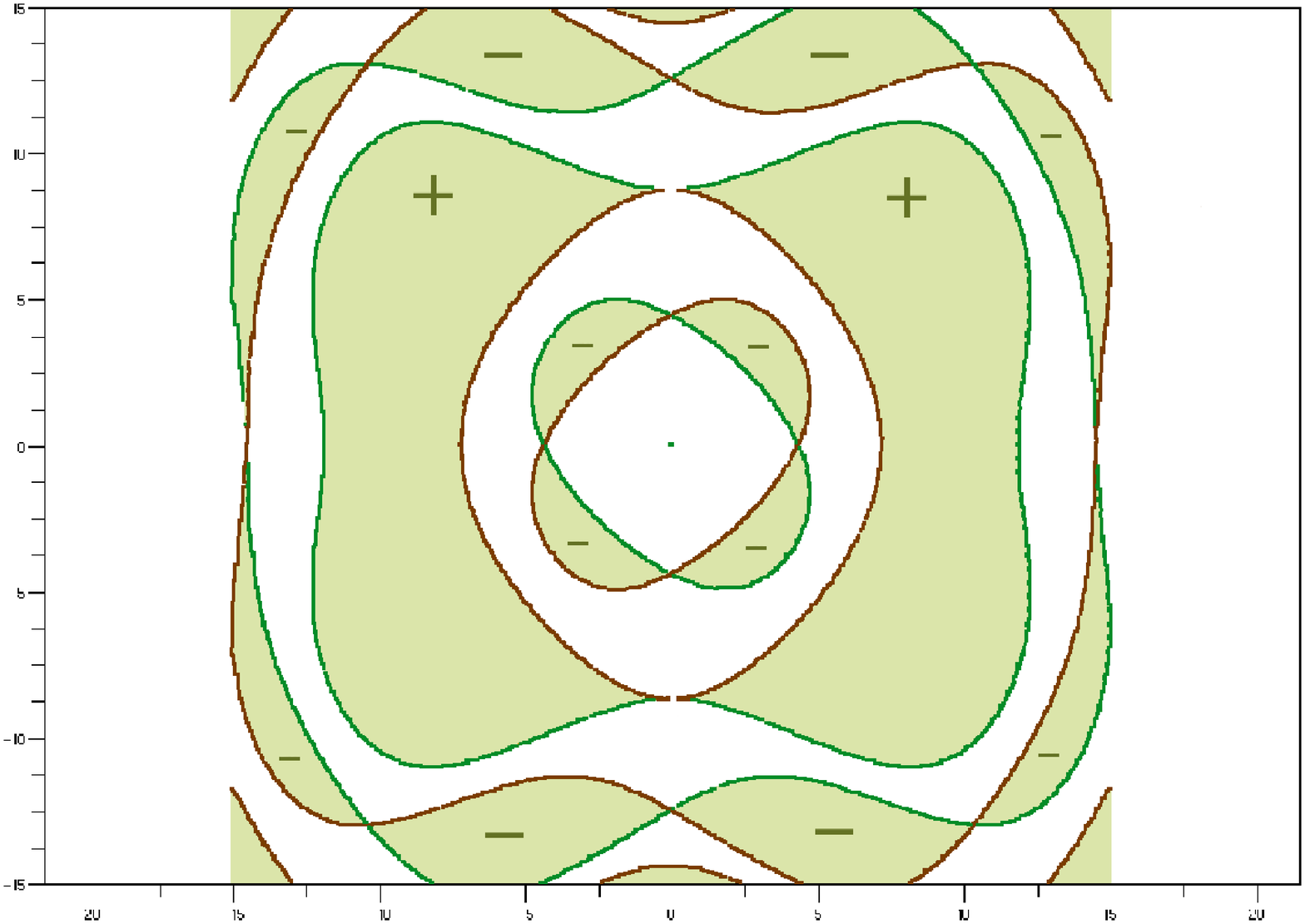}
 	\rput(-13.25,7.55){\Cafe{\Tiny $\bm{^{12.19}}$}}
 	\rput(-13.15,8.15){\Cafe{\Tiny $\bm{^{9.64}}$}}
\rput(-13.2,7){\Cafe{\Tiny $\bm{^{16.20}}$}}
 	\rput(-13,8.7){\Cafe{\Tiny $\bm{^{8.75}}$}}
\rput(-13.0,6.4){\Cafe{\Tiny $\bm{^{22.01}}$}}
 	\rput(-12.9,9.05){\Cafe{\Tiny $\bm{^{8.34}}$}}
\rput(-12.85,6){\Cafe{\Tiny $\bm{^{26.19}}$}}
	\rput(-12.7,9.4){\Cafe{\Tiny \rotatebox{-10}{$\bm{^{8.05}}$}}}
\rput(-12.7,5.7){\Cafe{\Tiny  \rotatebox{5}{$\bm{^{29.24}}$}}}
	\rput(-12.55,9.65){\Cafe{\Tiny \rotatebox{-12}{ $\bm{^{7.75}}$}}}
\rput(-12.5,5.43){\Cafe{\Tiny  \rotatebox{10}{$\bm{^{30.89}}$}}}
	\rput(-12.4,9.9){\Cafe{\Tiny \rotatebox{-14}{ $\bm{^{7.39}}$}}}
\rput(-12.35,5.2){\Cafe{\Tiny  \rotatebox{15}{$\bm{^{31.15}}$}}}
	\rput(-12.2,10.1){\Cafe{\Tiny \rotatebox{-16}{ $\bm{^{7.02}}$}}}
\rput(-12.2,5){\Cafe{\Tiny  \rotatebox{20}{$\bm{^{30.20}}$}}}
	\rput(-11.97,10.35){\Cafe{\Tiny \rotatebox{-25}{  $\bm{^{6.59}}$}}}
\rput(-11.97,4.8){\Cafe{\Tiny  \rotatebox{25}{$\bm{^{28.11}}$}}}
	\rput(-11.77,10.5){\Cafe{\Tiny  \rotatebox{-30}{$\bm{^{6.09}}$}}}
\rput(-11.77,4.6){\Cafe{\Tiny  \rotatebox{30}{$\bm{^{25.11}}$}}}
	\rput(-11.63,10.7){\Cafe{\Tiny  \rotatebox{-35}{$\bm{^{5.53}}$}}}
\rput(-11.63,4.4){\Cafe{\Tiny  \rotatebox{35}{$\bm{^{21.48}}$}}}
	\rput(-11.45,10.9){\Cafe{\Tiny  \rotatebox{-40}{$\bm{^{4.91}}$}}}
\rput(-11.45,4.25){\Cafe{\Tiny  \rotatebox{40}{$\bm{^{17.54}}$}}}
	\rput(-11.25,11){\Cafe{\Tiny \rotatebox{-50}{$\bm{^{4.24}}$}}}
\rput(-11.21,4.12){\Cafe{\Tiny \rotatebox{50}{$\bm{^{13.58}}$}}}
	\rput(-11.05,11.15){\Cafe{\Tiny \rotatebox{-55}{$\bm{^{3.51}}$}}}
\rput(-10.97,4.07){\Cafe{\Tiny \rotatebox{55}{$\bm{^{9.87}}$}}}
	\rput(-10.85,11.25){\Cafe{\Tiny \rotatebox{-60}{$\bm{^{2.76}}$}}}
\rput(-10.75,3.95){\Cafe{\Tiny \rotatebox{60}{$\bm{^{6.65}}$}}}
	\rput(-10.7,11.37){\Cafe{\Tiny \rotatebox{-70}{$\bm{^{2.00}}$}}}
\rput(-10.6,3.85){\Cafe{\Tiny \rotatebox{70}{$\bm{^{4.06}}$}}}
	\rput(-10.47,11.45){\Cafe{\Tiny \rotatebox{-80}{$\bm{^{1.33}}$}}}
\rput(-10.35,3.8){\Cafe{\Tiny \rotatebox{80}{$\bm{^{2.2}}$}}}
	\rput(-10.2,11.57){\Cafe{\Tiny \rotatebox{-90}{$\bm{^{0.07}}$}}}
 \rput(-10.1,3.75){\Cafe{\Tiny \rotatebox{90}{$\bm{^{0.1}}$}}}
	\rput(-6.5,7.55){\Cafe{\Tiny $\bm{^{12.19}}$}}
	\rput(-6.6,8.17){\Cafe{\Tiny $\bm{^{16.20}}$}}
\rput(-6.65,7.0){\Cafe{\Tiny $\bm{^{9.64}}$}}
	\rput(-6.8,8.75){\Cafe{\Tiny $\bm{^{22.01}}$}}
\rput(-6.85,6.45){\Cafe{\Tiny $\bm{^{8.75}}$}}
	\rput(-6.9,9.05){\Cafe{\Tiny $\bm{^{26.19}}$}}
\rput(-7.,6.1){\Cafe{\Tiny $\bm{^{8.34}}$}}
	\rput(-7.1,9.35){\Cafe{\Tiny  \rotatebox{10}{$\bm{^{29.24}}$}}}
\rput(-7.2,5.75){\Cafe{\Tiny \rotatebox{-10}{$\bm{^{8.05}}$}}}
 	\rput(-7.3,9.7){\Cafe{\Tiny  \rotatebox{15}{$\bm{^{30.89}}$}}}
\rput(-7.45,5.5){\Cafe{\Tiny\rotatebox{-15}{ $\bm{^{7.75}}$}}}
	\rput(-7.45,9.97){\Cafe{\Tiny \rotatebox{20}{$\bm{^{31.15}}$}}}
\rput(-7.6,5.3){\Cafe{\Tiny\rotatebox{-20}{ $\bm{^{7.39}}$}}}
	\rput(-7.61,10.17){\Cafe{\Tiny   \rotatebox{25}{$\bm{^{30.20}}$}}}
\rput(-7.7,5.05){\Cafe{\Tiny \rotatebox{-25}{$\bm{^{7.02}}$}}}
	\rput(-7.85,10.39){\Cafe{\Tiny \rotatebox{30}{$\bm{^{28.11}}$}}}
\rput(-7.93,4.8){\Cafe{\Tiny\rotatebox{-30}{ $\bm{^{6.59}}$}}}
	\rput(-8.,10.6){\Cafe{\Tiny \rotatebox{35}{$\bm{^{25.11}}$}}}
\rput(-8.15,4.6){\Cafe{\Tiny\rotatebox{-35}{ $\bm{^{6.09}}$}}}
	\rput(-8.2,10.75){\Cafe{\Tiny \rotatebox{40}{$\bm{^{21.48}}$}}}
\rput(-8.3,4.45){\Cafe{\Tiny\rotatebox{-40}{ $\bm{^{5.53}}$}}}
	\rput(-8.4,10.9){\Cafe{\Tiny \rotatebox{45}{$\bm{^{17.54}}$}}}
\rput(-8.5,4.35){\Cafe{\Tiny\rotatebox{-45}{ $\bm{^{4.91}}$}}}
\rput(-8.7,4.2){\Cafe{\Tiny \rotatebox{-50}{$\bm{^{4.24}}$}}}
	\rput(-8.6,11.1){\Cafe{\Tiny \rotatebox{50}{$\bm{^{13.58}}$}}}
\rput(-8.85,4.05){\Cafe{\Tiny \rotatebox{-60}{$\bm{^{3.51}}$}}}
	\rput(-8.8,11.2){\Cafe{\Tiny \rotatebox{60}{$\bm{^{9.87}}$}}}
\rput(-9.05,3.94){\Cafe{\Tiny \rotatebox{-70}{$\bm{^{2.76}}$}}}
	\rput(-9,11.3){\Cafe{\Tiny \rotatebox{70}{$\bm{^{6.65}}$}}}
\rput(-9.3,3.85){\Cafe{\Tiny \rotatebox{-80}{$\bm{^{2.00}}$}}}
	\rput(-9.2,11.4){\Cafe{\Tiny \rotatebox{80}{$\bm{^{4.06}}$}}}
\rput(-9.5,3.75){\Cafe{\Tiny \rotatebox{-90}{$\bm{^{1.33}}$}}}
\rput(-9.7,3.65){\Cafe{\Tiny \rotatebox{-90}{$\bm{^{0.07}}$}}}
	\rput(-12.28,8.34){\Verde{\Tiny $\bm{^{0.25}}$}}
	\rput(-12.25,8.65){\Verde{\Tiny $\bm{^{0.31}}$}}
	\rput(-12.28,8.15){\Verde{\Tiny $\bm{^{0.21}}$}}
	\rput(-12.18,8.9){\Verde{\Tiny $\bm{^{0.34}}$}}
	\rput(-12.25,7.8){\Verde{\Tiny \rotatebox{5}{$\bm{^{0.11}}$}}}
	\rput(-12.1,9.12){\Verde{\Tiny \rotatebox{-10}{$\bm{^{0.36}}$}}}

	\rput(-11.9,9.45){\Verde{\Tiny \rotatebox{-20}{$\bm{^{0.38}}$}}}
\rput(-11.35,7.6){\Verde{\Tiny \rotatebox{40}{$\bm{^{-0.06}}$}}}
	\rput(-11.7,9.65){\Verde{\Tiny \rotatebox{-30}{$\bm{^{0.39}}$}}}
\rput(-11.22,7.35){\Verde{\Tiny \rotatebox{40}{$\bm{^{-0.18}}$}}}
	\rput(-11.5,9.83){\Verde{\Tiny \rotatebox{-40}{$\bm{^{0.38}}$}}}
\rput(-11.05,7.15){\Verde{\Tiny \rotatebox{40}{$\bm{^{-0.27}}$}}}
	\rput(-11.3,9.98){\Verde{\Tiny \rotatebox{-60}{$\bm{^{0.37}}$}}}
\rput(-10.9,6.95){\Verde{\Tiny \rotatebox{40}{$\bm{^{-0.33}}$}}}
	\rput(-11.05,10.1){\Verde{\Tiny \rotatebox{-70}{$\bm{^{0.35}}$}}}
\rput(-10.7,6.78){\Verde{\Tiny \rotatebox{40}{$\bm{^{-0.36}}$}}}
	\rput(-10.75,10.07){\Verde{\Tiny \rotatebox{-80}{$\bm{^{0.31}}$}}}
\rput(-10.57,6.61){\Verde{\Tiny \rotatebox{40}{$\bm{^{-0.36}}$}}}
	\rput(-10.5,10.07){\Verde{\Tiny \rotatebox{-95}{$\bm{^{0.26}}$}}}
\rput(-10.38,6.52){\Verde{\Tiny \rotatebox{40}{$\bm{^{-0.33}}$}}}
	\rput(-10.28,10.02){\Verde{\Tiny \rotatebox{-105}{$\bm{^{0.20}}$}}}
\rput(-10.20,6.37){\Verde{\Tiny \rotatebox{40}{$\bm{^{-0.28}}$}}}
 	\rput(-10,10.){\Verde{\Tiny \rotatebox{-115}{$\bm{^{0.12}}$}}}
\rput(-10.05,6.24){\Verde{\Tiny \rotatebox{40}{$\bm{^{-0.21}}$}}}
	\rput(-9.75,9.9){\Verde{\Tiny \rotatebox{-130}{$\bm{^{0.04}}$}}}
\rput(-9.73,8.97){\Verde{\Tiny \rotatebox{40}{$\bm{^{-0.21}}$}}}
\rput(-9.55,8.83){\Verde{\Tiny \rotatebox{40}{$\bm{^{-0.28}}$}}}
\rput(-9.38,8.7){\Verde{\Tiny \rotatebox{40}{$\bm{^{-0.33}}$}}}
\rput(-9.25,8.55){\Verde{\Tiny \rotatebox{40}{$\bm{^{-0.36}}$}}}
\rput(-9.10,8.39){\Verde{\Tiny \rotatebox{40}{$\bm{^{-0.36}}$}}}
\rput(-8.95,8.25){\Verde{\Tiny \rotatebox{40}{$\bm{^{-0.33}}$}}}
\rput(-8.8,8.05){\Verde{\Tiny \rotatebox{40}{$\bm{^{-0.27}}$}}}
\rput(-8.6,7.86){\Verde{\Tiny \rotatebox{40}{$\bm{^{-0.18}}$}}}
\rput(-8.5,7.6){\Verde{\Tiny \rotatebox{40}{$\bm{^{-0.06}}$}}}

\rput(-7.6,7.3){\Verde{\Tiny $\bm{^{0.11}}$}}
\rput(-7.55,7.){\Verde{\Tiny $\bm{^{0.21}}$}}
\rput(-7.53,6.8){\Verde{\Tiny \rotatebox{-5}{$\bm{^{0.25}}$}}}
\rput(-7.55,6.5){\Verde{\Tiny \rotatebox{-7.5}{$\bm{^{0.31}}$}}}
\rput(-7.75,6.25){\Verde{\Tiny \rotatebox{-10}{$\bm{^{0.34}}$}}}
\rput(-7.8,6.05){\Verde{\Tiny \rotatebox{-20}{$\bm{^{0.36}}$}}}
\rput(-8.0,5.75){\Verde{\Tiny \rotatebox{-30}{$\bm{^{0.38}}$}}}
\rput(-8.2,5.6){\Verde{\Tiny \rotatebox{-40}{$\bm{^{0.39}}$}}}
\rput(-8.5,5.4){\Verde{\Tiny \rotatebox{-50}{$\bm{^{0.38}}$}}}
\rput(-8.7,5.25){\Verde{\Tiny \rotatebox{-60}{$\bm{^{0.37}}$}}}
\rput(-8.9,5.17){\Verde{\Tiny \rotatebox{-70}{$\bm{^{0.35}}$}}}
\rput(-9.15,5.15){\Verde{\Tiny \rotatebox{-80}{$\bm{^{0.31}}$}}}
\rput(-9.4,5.15){\Verde{\Tiny \rotatebox{-90}{$\bm{^{0.26}}$}}}
\rput(-9.6,5.17){\Verde{\Tiny \rotatebox{-100}{$\bm{^{0.20}}$}}}
\rput(-9.8,5.25){\Verde{\Tiny \rotatebox{-110}{$\bm{^{0.12}}$}}}
\rput(-10.,5.4){\Verde{\Tiny \rotatebox{-110}{$\bm{^{0.04}}$}}}

\psset{linecolor=green}
\dotnode[dotstyle=*,dotscale=1.1 1.1,linecolor=cafe](-15.73,13.4){p1}
\Rput[tl](-13.61,15.4){\rnode{m1}{\tiny$\left\|\begin{array}{cc}
              1.0006 & 0.0001 \\7.4113 & 1.0006
             \end{array}\right\|$}}
\dotnode[dotstyle=*,dotscale=1.1 1.1,linecolor=cafe](-15.58,10.8){p2}
\Rput[tl](-19.1,11.5){\rnode{m2}{\tiny$\left\|\begin{array}{cc}
              -0.9991 & 0.0004 \\-4.3269 & -0.9991
             \end{array}\right\|$}}
\dotnode[dotstyle=*,dotscale=1.1 1.1](-14.05,12.09){p3}
\Rput[tl](-19.61,15.4){\rnode{m3}{\tiny$\left\|\begin{array}{cc}
              1.0000 & 0.8251 \\0.0002 & 1.0000
             \end{array}\right\|$}}
\dotnode[dotstyle=*,dotscale=1.1 1.1](-14.9,7.47){p4}
\Rput[tl](-19.1,8.25){\rnode{m4}{\tiny$\left\|\begin{array}{cc}
              0.9998 & 0.4451 \\-0.0005 & 0.9998
             \end{array}\right\|$}}
\dotnode[dotstyle=*,dotscale=1.1 1.1](-6.9,2.5){p6}
\Rput[tl](-9.86,1){\rnode{m6}{\tiny$\left\|\begin{array}{cc}
              -1.0000 & 0.7673 \\0.0001 & -1.0000
             \end{array}\right\|$}}
\dotnode[dotstyle=*,dotscale=1.1 1.1](-3.65,5.5){p7}
\Rput[tl](-3.41,6.15){\rnode{m7}{\tiny$\left\|\begin{array}{cc}
              -1.0000 & -0.1235 \\-0.0000 & -1.0000
             \end{array}\right\|$}}
\dotnode[dotstyle=*,dotscale=1.1 1.1](-7.65,3.3){p8}
\Rput[tl](-3.01,3.9){\rnode{m8}{\tiny$\left\|\begin{array}{cc}
              1.0001 & 1.1148 \\0.0002 & 1.0001
             \end{array}\right\|$}}
\rput(-2,9.55){\Verde{\scriptsize {$\bm{{B}}$}}}
\rput(-9.5,12){\blue{\scriptsize {$\bm{{A}}$}}}
\dotnode[linecolor=blue,dotstyle=*,dotscale=1.1 1.1](-9.5,11.75){A}
\dotnode[dotstyle=*,dotscale=1.1 1.1](-8.85,8.7){p9}
\Rput[tl](-3.41,9.5){\rnode{m9}{\tiny$\left\|\begin{array}{cc}
              -1.0002 & 0.3688 \\0.0013 & -1.0002
             \end{array}\right\|$}}
\dotnode[dotstyle=*,dotscale=1.1 1.1,linecolor=cafe](-6.9,12.75){p12}
\Rput[tl](-4.11,15.4){\rnode{m12}{\tiny$\left\|\begin{array}{cc}
              -1.0000 & -0.0000 \\-11.8416 & -1.0000
             \end{array}\right\|$}}
 \dotnode[dotstyle=*,dotscale=1.4 1.4](-15.87,5.65){p14}
 \Rput[tl](-19.1,3.75){\rnode{m14}{\tiny$\left\|\begin{array}{cc}
               -0.9999 & 0.1325 \\-0.0000 &-0.9999
              \end{array}\right\|$}}

\ncangle[linecolor=USgray,angleA=90,angleB=270]{->}{p1}{m1}
\ncangle[linecolor=USgray,angleA=180,angleB=0]{->}{p2}{m2}
\ncangle[linecolor=USgray,angleA=180,angleB=0]{->}{p3}{m3}
\ncangle[linecolor=USgray,angleA=180,angleB=0]{->}{p4}{m4}
\ncangle[linecolor=USgray,angleA=0,angleB=180]{->}{p5}{m5}
\ncangle[linecolor=USgray,angleA=180,angleB=90]{->}{p6}{m6}
\ncline[linecolor=USgray,angleA=90,angleB=90]{->}{p7}{m7}
\ncline[linecolor=USgray,angleA=0,angleB=180]{->}{p8}{m8}
\ncangle[linecolor=USgray,angleA=0,angleB=180]{->}{p9}{m9}
\ncangle[linecolor=USgray,angleA=180,angleB=0]{->}{p10}{m10}
\ncangle[linecolor=USgray,angleA=0,angleB=180]{->}{p11}{m11}
\ncangle[linecolor=USgray,angleA=0,angleB=270]{->}{p12}{m12}
\ncangle[linecolor=USgray,angleA=90,angleB=-90]{p13}{m13}
\ncangle[linecolor=USgray,angleA=180,angleB=0]{->}{p14}{m14}

 \end{pspicture}
\caption{\label{fig:bicafeverde}\small The map of the stability, resonance areas and the separatrix borders of types $(+)$ and $(-)$ on the intensity map $\beta_1, \beta_2$, scanned by integrating the matrix equation (\ref{eqforb}) for the biharmonic $b(1)$ in the periodicity interval $[0,1]$. The separatrix borderlines host the exceptional matrices of types (\ref{kickdeltas}), (\ref{ODD}). The branches yielding the oscillator kicks are brown, and the ones representing the distorted free evolution are green. The approximate values of the distorted evolution time $\tau$ and the simulated oscillator kicks $a$ are given by the sequences of numbers along the corresponding borderlines. Notice the negative $\tau$ on two sections of the internal green branch.
 The points $A$, $B$, marked blue and green represent the field amplitudes generating the space trajectories of Fig.~6 and Fig.~8 respectively. 
}
\end{sidewaysfigure}

 With some surprise we found that (\ref{kickdeltas}) and/or (\ref{ODD}) occur on the separatrix branches of Fig.~\ref{fig:bicafeverde} with either $\mathrm{Tr}\,b(1)=+2$ (we call them of type $+$), or $\mathrm{Tr}\,b(1)=-2$ (type $-$). The result contained some mystery. Though it is quite obvious that for $|\mathrm{Tr}b(1)|=2$, the matrix $b(1)$ degenerates, this still does not imply that it must appear in the standard Jordan's form (\ref{kickdeltas}) or (\ref{ODD}) precisely in the $q, p$ basis. 
 
As turns out, the phenomenon occurs always when $\beta(t)^2$ in eq. (\ref{eqforb}) is symmetric with respect to the center of the operation interval. To show this it is convenient to fix the time variable $t$ with $t=0$ at the symmetry center; then to consider the evolution matrix $b(t)=b(t,-t)$ generated in the expanding interval $[-t,t]$ (see \cite{MRPS}). 

\begin{proposition}\label{propo:4}
Suppose $\Lambda(t)$ is symmetric around the point $t=0$, $i.e.$ $\Lambda(-t)\equiv\Lambda(t)$. Then if at any $t\in\mathbb{R}$, $\mathrm{Tr}\,b(t)=2$, the matrix $b(t)=b(t,-t)$ must take one of the forms (\ref{kickdeltas}), but if  $\mathrm{Tr}\,b(t)=-2$, then $b(t)$ must take one of the forms (\ref{ODD}).
\end{proposition}
\textit{Proof.} For the symmetric $\Lambda(t)$ given by (\ref{eqforb}) the evolution matrix $b=b(t,-t)$ fulfills:
\begin{eqnarray}
 \frac{db}{dt}=\Lambda(t)b+b\Lambda(t)\label{70}
	=(b_{21}-\beta^2b_{12})\mathbb{1}+\big(\mathrm{Tr}\,b\,\big)\Lambda(t)
\end{eqnarray}
Hence one easily shows:
\begin{eqnarray}
 \frac{d}{dt}\left[ 
b_{12}b_{21}-\frac{1}{4}(\mathrm{Tr}\,b)^2\label{73}
\right]\equiv0\Rightarrow b_{12}b_{21}-\frac{1}{4}(\mathrm{Tr}\,b)^2=C=const.
\end{eqnarray}
The constant in (\ref{73}) can be determined by shrinking the interval $[-t,t]$ to zero. For $t=0$, the evolution matrix is trivial, $b=b(0,0)=\mathbb{1}, b_{12}=b_{21}=0$, $i.e.$, the constant in (\ref{73}) is $C=-\frac{1}{4}(\mathrm{Tr}\,b)^2=-1$, implying
\begin{eqnarray}
 b_{12}b_{21}=\frac{1}{4}(\mathrm{Tr}\,b)^2-1. \label{b12b21}
\end{eqnarray}
Hence, $b_{12}b_{21}$ must vanish whenever $\mathrm{Tr}\,b=\pm2$. So, either $b_{12}$ or $b_{21}$ must vanish, leading to one of Jordan's forms (\ref{kickdeltas}) or (\ref{ODD}) on borders of type $(+)$ or $(-)$.$\fullsquare$\\

\textit{Corollary}. While under our symmetry assumptions the values $b_{12}=0$ and $b_{21}=0$ appear on the stability thresholds, none of them can occur inside of the squeezing areas. Indeed, (\ref{b12b21}) imply that for $|Tr\;b|>2$ none of $b_{12}$ and $b_{21}$ can vanish.\\

Note that the Proposition 4 applies to the biharmonic $\beta(t)$, odd with respect to the centers of the periodicity intervals $[n, n+1]$, and vanishing at the extremes $n=0,\pm1,\pm2,\cdots$\footnote{For the harmonic $\beta(t)$, the assumptions of Proposition \ref{propo:4} are satisfied in the periodicity intervals $[n-\frac{1}{2}, n+\frac{1}{2}]$ thanks to the symmetry of $\beta (t)$. However, due to the existence of the constant field component, the vector potential $A$re does not vanish at the operation extremes, affecting the physical sense of (\ref{kickdeltas}) and (\ref{ODD}) (c.f. the remarks in Dodonov et al. \cite{Dodonov:1994}).}. The corresponding stability borders (Fig.~\ref{fig:bicafeverde}) form a sequence of closed, intersecting pairs of branches surrounding the map center. The computer scanning shows that the inner pair, intersecting at 4 points, is of type $(-)$, $i.e.$, hosts the evolution matrices (\ref{ODD}). The next pair, with 2 intersections, is of type $(+)$, hosting (\ref{kickdeltas}). In general, the $1^{st}$, $3^{rd}$, and all odd pairs are of type $(-)$, whereas all even pairs are of type $(+)$ (see Fig. \ref{fig:bicafeverde}). On each pair, one branch (brown) offers the soft imitations of the $\delta$-kicks (with the amplitudes $a\neq 0$, but $\tau=0$), while the other one (green), yields the incidents of the distorted free evolution ($a=0$, $\tau\neq 1$). Note, that the operations on the $(+)$ branches give the clean effects after the single pulse period (or its multiples), while the operations of the $(-)$ branches yield the parity free effects $2a$ or $2\tau$ after two pulse periods (or $2na$, $2n\tau$ after $2n$  repetitions). To illustrate the details, our Fig. \ref{fig:bicafeverde} reports the values of the ``distorted time" $\tau$ along the green branches and the amplitudes of the effective oscillator kicks along the first brown branch.

\begin{figure}[t]
\centering
{
\includegraphics[width=0.52\linewidth]{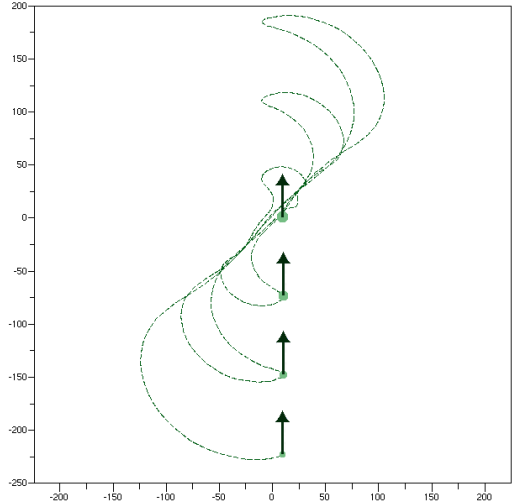}}
\begin{quote}
\caption{\label{fig:inversion}\small The double period operation for the biharmonic field amplitudes $(\beta_1,\beta_2)=(2.40,2.68)$ on the first separatrix branch. Under the repeated applications of the $2$ period pulse pattern, the Gaussian packet shrinks instead of expanding and its center travels in the direction opposite to its initial velocity, simulating the incidents of the inverted free evolution with $\Delta t=n\tau\cong -0.3688n$ at $n= 2,4,6\cdots$. }
\end{quote}
\end{figure}

Our computer scanning shows that the double period operations of the type $(-)$ have the exactly vanishing Floquet centers, and so, produce the stable (resistant) effects, while the operations of type $(+)$ define non-vanishing fuzzy centers and can be affected by constant external forces.

%
%

As an example, we have choosen a case of the free evolution inversion for a point on the first (negative) separatrix branch of Fig. \ref{fig:bicafeverde}. The effects of the ``retrospective operations'' after $4$ double periods are illustrated by the center of a wave packet recovering in the time instants $t=2n$ $(n=1, 2,\cdots)$ its past shapes (shrinking instead of spreading in some number of steps); its subsequent positions showing a sequence of shifts in the direction opposite to the initial velocity (Fig. \ref{fig:inversion}). 

In turn, the possibility of simulating the $\delta(t)$ pulses of the oscillator forces by the soft biharmonic fields (brown branches of Fig.~\ref{fig:bicafeverde}) is no less essential: it saves the realistic sense of all operations programmed with the help of the sharp oscillator pulses \cite{David:1994,Bogdan:1990,Dodonov:1993}.

\section{Dark areas: the squeezing}
 Below we shall not comment any particular \textit{squeezed states}, but rather the \textit{squeezing operator} which can deform any wave packet, no matter, is it Gaussian or not, and we shall be looking for the origin of this phenomenon in the structure of the full period evolution (Floquet) operator $U(1)$ and its matrix $u(1)$.

The squeezing mechanisms were carefully examined in an ample sequence of papers, the most complete ones by Dodonov et al \cite{Dodonov:1994,Yuen:1976,Ma:1989,Royer:1985,Brown:1979,Dodonov:2002,Castanos:1994,Dodonov:1993,MRPS,Ba,David:1990,Adoc,DM,DMM72}; so our approach cannot dissent too much, though it can contribute with some more observations:\\

 The use of the quadratic Hamiltonians with the time dependent $B=B(t)$ is justified as the first step of the known EIH method \cite{EIH} for not too ample experiment areas.\\

 In case of periodic $\beta(t)\equiv\beta(t+1)$ in the cylindrical geometry the information about the squeezing can be read from the $2\times2$ oscillatory matrix $b(1)$, even if the time dependent field pulses do not distinguish any invariant ``squeezing center''. This holds also if the magnetic field vanishes at the beginning and at the end of the operation. If $| \mathrm{Tr} \,b(1)|>2$ (dark areas), then $\lambda_{\pm}$ are real, $\lambda_{+} \lambda_{-}=1$, describing a squeezing on the canonical plane. For $\mathrm{Tr}\,b(1)>2$ (type +), the transformation is a squeezing \textit{sensu stricto} with two positive coefficients $\lambda_{\pm}>0$, but if $\mathrm{Tr}\,b(1)<-2$ (type -), then $b(1)$ yields the superposition of squeezing and the parity transformation (see also our recent report \cite{MRPS}).  The squeezed or amplified canonical observables are defined just by the (row) eigenvectors of $b(1)$ (c.f. \cite{MRPS}).\\

Some studies dedicate special attention to the \textit{scale squeezing} or expansion in which $q$ is squeezed at the cost of $p$ or \textit{vice versa} \cite{Brown:1979,DM}. Here, the cases of $b_{12}=0$ or $b_{21}=0$ are quite relevant not only on the separatrix branches,  but also inside of the squeezing areas. The value $b_{12}=0$ in $b(1)$ implies the coordinate $q$ squeezed or expanded at the cost of some other canonical observable, while $b_{21} = 0$, the same for the canonical momentum. The coincidence $b_{12} = b_{21} =0$, would represent  the position $q$ (meaning $x$ and/or $y$) squeezed or amplified exactly at the cost of the corresponding momentum ($p_x$ or $p_y$). However, due to our Proposition 4, none of these cases can happen if $\beta^2$ is symmetric around the operation center, when the desired zeroes of $b_{12}$ and/or $b_{21}$ are all located on the separatrix, without penetrating ever inside the dark areas. Hence, the scale transformations can be achieved only in the operation intervals with an asymmetric $\beta(t)^2$. In this last case the role of the `fuzzy points' again calls attention. If the Floquet center $\vect{X}$ of the classical/quantum trajectory vanishes, then the squeezing operations will be immune to the constant external force $F$. However, should $\vect{X}$ be fuzzy, then if the field pulses are repeated, the $q$-trajectory, apart of the sequence of the squeezing operations, will show a drift in a direction orthogonal to F. So, would the ``fuzzy center'' be indeed a generalized element of noncommutative geometry applicable for variable fields?

\section{Open Problems.} 

Since the effects which we discussed are computed in the dimensionless field variables, it is not excluded that their analogues can  reappear in different orders of magnitudes as $e.g.$ interiors of the oscillating cristal lattices, the nodal points of the crossed electromagnetic waves \cite{PhysRevLett.59.1659,Bogdan:1990}, which do not involve the macroscopic traps \cite{RevModPhys.58.699}. Supposing that some analogues `in little' can exist, they might contribute to different topics recently studied.

The behavior of particle gas in $2D$ in presence of the harmonic or biharmonic fields, shows a sequence of surprising reactions to the external electric force, in form of a ``giant'' \cite{nobels:2007} or vanishing, or even negative resistance \cite{Nature:2002}. The effects are apparently due to the presence of material shells forming a sandwich on both sides of the motion plane; but the theoretical discussions are not yet concluded (see e.g. Zudov et al. \cite{Zudov:2006}). One might henceforth ask, whether the surprising phenomena have not some elementary counterparts for time dependent fields? Could e.g., the behavior of the $2D$ evolution loops with the vanishing `fuzzy center',  be interpreted as an analogue of an `anomalous resistance'?  Could the distorted free evolution mean a modified effective  time? Could the squeezing effects matter? Last not least, could the analogous phenomena affect the polarized vacuum, if such an entity exists \cite{SqVa}?

\section*{Acknowledgments}
The authors are indebted for the interest of their colleagues in the Physics and Mathematics Departments of Cinvestav, M\'exico. One of us (BM) acknowledges the support of the Conacyt project 49253-F. The technical assistance of Eng. Erasmo G\'omez is gratefully acknowledged.

\newpage
\section*{Bibliography}
\bibliography{fuzzy.bib}


\end{document}